\documentclass[lettersize,journal,onecolumn]{IEEEtran}
\usepackage{amsmath,amsfonts}
\usepackage{algorithmic}
\usepackage{algorithm}
\usepackage{array}
\usepackage[caption=false,font=normalsize,labelfont=sf,textfont=sf]{subfig}
\usepackage{textcomp}
\usepackage{stfloats}
\usepackage{url}
\usepackage{verbatim}
\usepackage{graphicx}
\usepackage{cite}
\begin{document}

\title{Image-aware Layout Generation\\ with User Constraints for Poster Design}

\author{Chenchen Xu, Kaixin Han, and Weiwei Xu
        % <-this % stops a space
\thanks{Chenchen Xu and Weiwei Xu are with the State Key Lab of CAD\&CG, Zhejiang University, China. Kaixin Han is with the College of Computer Science and Technology, Zhejiang University, China.}
\thanks{E-mail: xuchenchen@zju.edu.cn, hankx@zju.edu.cn, and xww@cad.zju.edu.cn}
}

\markboth{Image-aware Layout Generation with User Constraints for Poster Design}%
{Shell \MakeLowercase{\textit{et al.}}: A Sample Article Using IEEEtran.cls for IEEE Journals}


\maketitle

\begin{abstract}
Graphic layout is essential in poster generation. Professionals often need to design different layouts for a product image, to ensure they meet specific user requirements. This paper focuses on utilizing a deep-learning model to automatically generate image-aware layouts with user-defined constraints, including layout attributes and partial layouts. Layout attribute constraints require generated layouts to include and exclude elements of specified classes, such as text, logos, underlays, and embellishments. Our model represents different attributes by sampling multidimensional Gaussian noise with different means, and we propose an attribute-consistent loss and an attribute-disentangled loss to ensure that the generated layout satisfies the specified attribute. Partial layout constraints provide our model with incomplete layout information to guide the generation of the remaining elements. We design a partial-constraint loss to incorporate the provided partial layout. Furthermore, we introduce a random mask to diversify the partial layout constraints, which can encourage the model to learn more general latent representations of the provided partial layouts. Both quantitative and qualitative evaluations demonstrate that our model can generate different image-aware layouts according to various user constraints while achieving state-of-the-art performance.
\end{abstract}


\begin{figure*}[ht!]
\centering
\includegraphics[width=\textwidth]{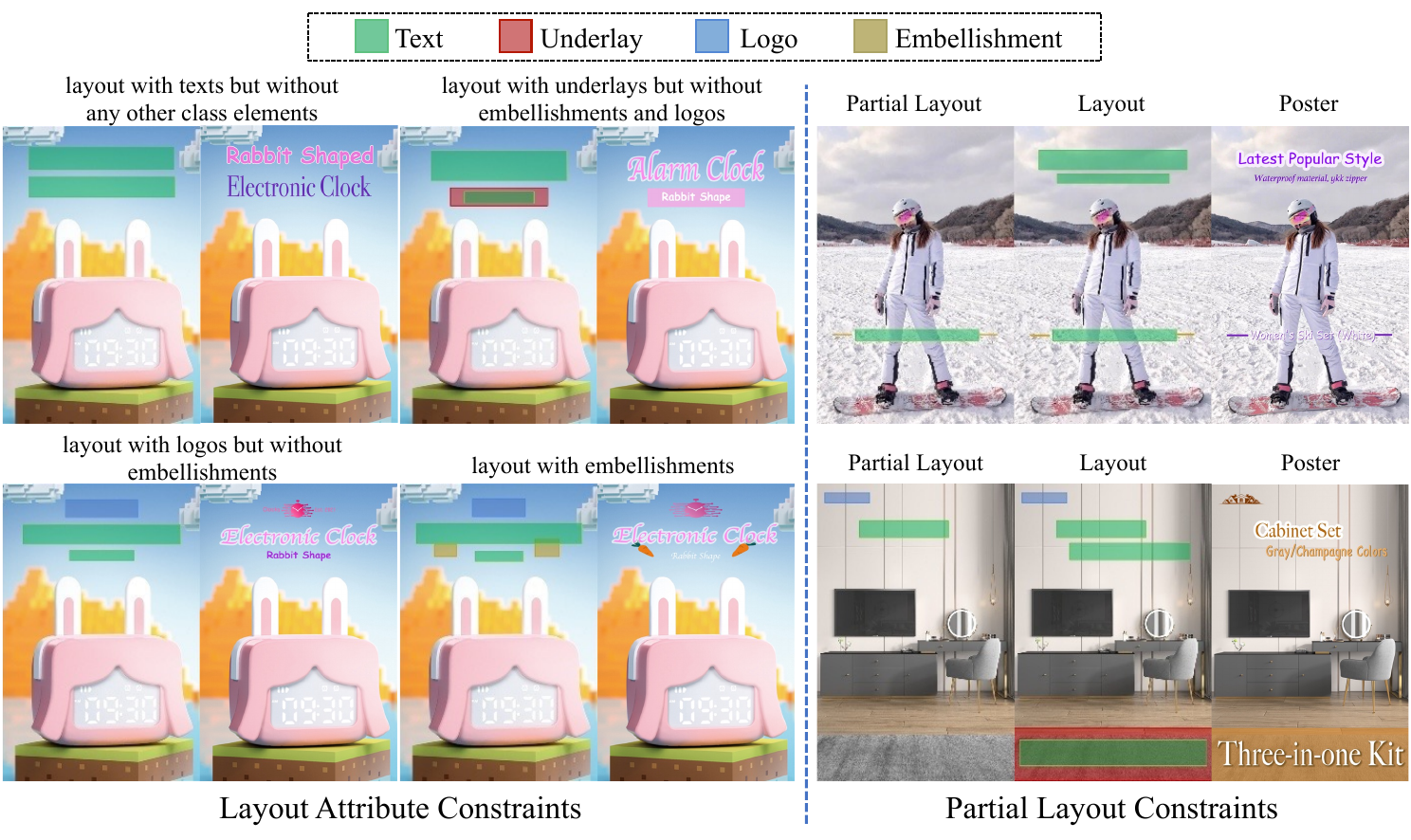}
\caption{{\bf Examples of generated layouts and posters with image contents and user constraints.} Our model generates image-aware layouts that adhere to layout attribute constraints (left) and partial layout constraints (right), which can be used to generate advertising posters.}
\label{fig:introduction}
\end{figure*}

\section{Introduction}
\IEEEPARstart{G}{raphic} layout design, which involves arranging texts, logos, underlays, and other 2D elements~\cite{DBLP:journals/pami/LiYHZX21,DBLP:conf/ijcai/ZhouXMGJX22}, is an essential component for various media, such as magazines~\cite{DBLP:journals/tip/KanungoM03,DBLP:conf/iui/SchrierDJWS08,DBLP:journals/tip/HedjamNKC15,DBLP:journals/tomccap/YangMXRL16,DBLP:conf/siggraph/TabataYMY19}, posters~\cite{qiang2016learning,qiang2019learning,DBLP:journals/jzusc/YouJYYS20,DBLP:conf/chi/GuoJSLL0C21}, web pages~\cite{DBLP:journals/tog/PangCLC16,DBLP:conf/mm/ZhangHRYXH17,DBLP:journals/india,liang2023sketch2wireframe}, and comics~\cite{calic2007efficient,cohn2013navigating,wang2021interactive,qiao2023design}. 
% An advertising poster designed with a high-quality layout not only looks visually pleasing but also communicates information clearly. 
The well-crafted graphic layouts are heavily reliant on the designers' experience and proficiency.

In the past decade, deep-learning-based methods for graphic-layout generation have emerge~\cite{DBLP:conf/iccv/JyothiDHSM19,DBLP:conf/cvpr/ArroyoPT21,DBLP:conf/iccv/GuptaLA0MS21,DBLP:journals/corr/abs-2303-05049}. Recently, some image-aware methods have been proposed to model the relationship between image content and graphic layout elements~\cite{DBLP:conf/ijcai/ZhouXMGJX22,DBLP:conf/mm/CaoMZLXGJ22,DBLP:conf/cvpr/XuZGJX23,DBLP:journals/corr/abs-2303-15937,li2023relation,horita2024retrieval}. However, these models are not well designed to handle user constraints that express diverse design demands. In this paper, we focus on four classes of elements: text, underlay, logo, and embellishment, and divide user constraints into two main categories: \emph{layout attribute} and \emph{partial layout} constraints. Layout attribute constraints are used to control that the generated layout includes the elements of required classes (attribute elements) and excludes the elements of undesired classes (undesired elements). For example, when the attribute is "layout with logos but without embellishments", generated layouts need to display the product logo without any embellishments. Partial layout constraints require the model to supplement the given incomplete layout and generate a complete layout. Although CGL-GAN~\cite{DBLP:conf/ijcai/ZhouXMGJX22} and PDA-GAN~\cite{DBLP:conf/cvpr/XuZGJX23} allow to guide the layout generation with user-specified coordinates and classes of partial elements, they do not always conform to such constraints and fail to handle the constraints with \emph{incomplete element information}, for instance, coordinate or class only. More importantly, these models cannot handle layout attribute constraints.

This paper focuses on generating different high-quality graphic layouts according to user constraints for one product image. To this end, our proposed network integrates the layout attribute and partial layout constraints into image-aware layout generation methods, abbreviated as the IUC-Layout network. As a result, it is controllable, enabling the designer to express the diverse presentation requirements of advertising posters in the layout generation. Our model samples multi-dimensional Gaussian noise with different means to represent different types of layout attribute constraints. Since this representation assigns each layout attribute constraint with a region, it can fill the empty region between different means with more training examples and force the network to learn the intrinsic representation of different attributes robust to the noise perturbation. We found that it is beneficial to improve the robustness of our model during training. Specifically, we sample 4 dimensional Gaussian noise to represent 4 types of layout attribute constraints. In addition, we design attribute-consistent loss and attribute-disentangled loss to ensure the layout generated by the IUC-Layout network satisfies the corresponding layout attribute constraint. They are achieved by approximately counting the number of attribute or undesired elements using softmax operation to facilitate the gradient backpropagation. As shown in Fig.~\ref{fig:introduction}, the graphic layout design by the model can arrange four classes of elements, including texts, underlays, logos, and embellishments, at the appropriate positions based on product images and user-specified layout attributes or partial layouts. When the layout attribute constraint is "layout with text but without any other class elements", as shown in the top-left of Fig.~\ref{fig:introduction}, the generated layout consists of text elements only.

Additionally, we introduce a random mask operation to obtain incomplete element information constraints for the partial layout, which can encourage our model to learn more general latent representations of provided partial layouts. We also propose a partial-constraint loss to guide models to generate layouts that are precisely consistent with the given information. In our experiments, we also integrate the partial-constraint loss and the random mask into other image-aware layout networks to further verify the benefits of these two operations when handling partial layout constraints.

%Overall, both quantitative and qualitative evaluations demonstrate that our model can generate different layouts for the same product image according to various user constraints, while also achieving state-of-the-art (SOTA) performances.
% In the experiments section, we will demonstrate the effectiveness of the partial-constraint loss on CGL-GAN \cite{DBLP:conf/ijcai/ZhouXMGJX22} and PDA-GAN \cite{DBLP:conf/cvpr/XuZGJX23}. 

% In addition, as mentioned earlier, many experiments and observations indicate that existing image-aware layout generation models ~\cite{DBLP:conf/ijcai/ZhouXMGJX22, DBLP:conf/cvpr/XuZGJX23} cannot fully satisfy partial layout constraints. Therefore, we propose a partial-constraint loss for these models to address this problem. Quantitative and qualitative experiments demonstrate that the model with constraint loss achieves optimal effect in user-constraint consistency. To increase the variety of user constraints such as different element categories, quantities, and coordinates, we introduce a random mask to the input. After training, our model can adapt to various user constraints. The proposed user constraint loss and constraint random mask can be easily transferred to other models. In the subsequent experiments, we will demonstrate the effectiveness of user constraint loss function on CGL-GAN \cite{DBLP:conf/ijcai/ZhouXMGJX22} and PDA-GAN \cite{DBLP:conf/cvpr/XuZGJX23}.

We summarize the contributions of this paper as follows:
\begin{itemize}
\item we design an efficient representation of layout attribute constraints, which can force the network to learn the intrinsic representation of different attributes robust to the noise perturbation. Two losses, attribute-consistent loss and attribute-disentangled loss, are designed to ensure that the generated layouts by IUC-Layout network satisfy user-specified attributes.
    
\item We design a partial-layout loss that guides the model to complete layouts based on the given information. Furthermore, we introduce a random mask operation to obtain incomplete element information constraints for the partial layout, to enhance the model's general latent representations.
    
\item Both quantitative and qualitative evaluations demonstrate that our model can generate different high-quality layouts according to one product image with various user constraints while achieving SOTA performances.
\end{itemize}

\begin{figure*}[t]
\centering
\hspace{0cm}\includegraphics[width=\textwidth]{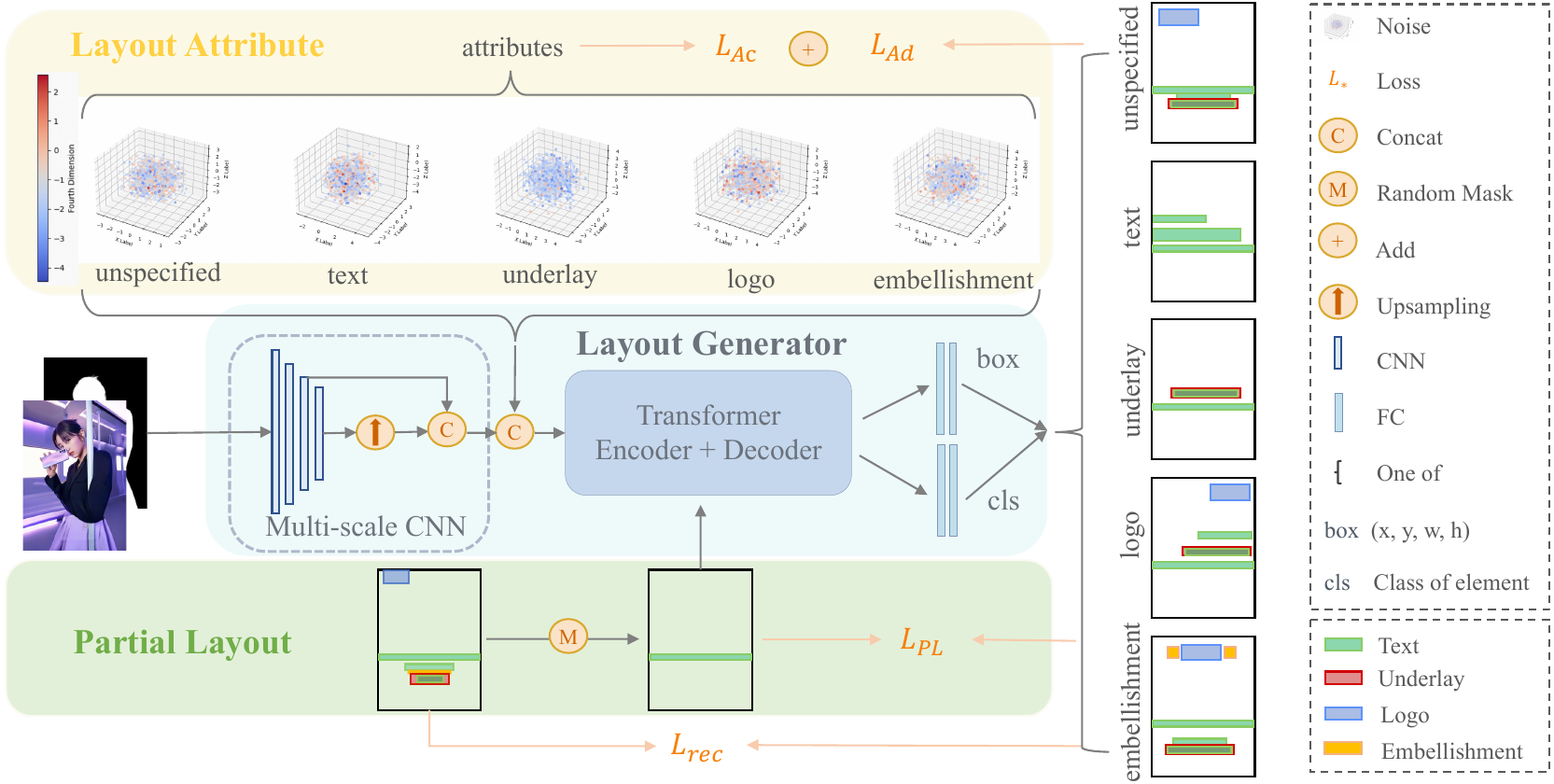}
\caption{{\bf The architecture of our network.} 
The three-dimensional views along with the color map visualize the sampled 4-dimensional Gaussian noise. During each training step, our model samples noise according to the specified attribute and combines it with image contents and the partial layout to generate an image-aware layout that satisfies user constraints.}
\label{fig:model}
\end{figure*}

\begin{table*}[!t]
    \centering
    \setlength{\tabcolsep}{1.8mm}
    \renewcommand{\arraystretch}{1.5}
    \caption{$a$ and $\mathcal{S}^*$ respectively represent the attribute element and the sets of undesired elements for the specified attribute $*$.}

    {
    \scalebox{1.1}{
    \begin{tabular}{c|c|c}
    \hline
         \bf{layout attribute ($*$)}    &\bf{attribute element ($a$)}  &\bf{set of undesired classes  ($\mathcal{S}^*$)}\\
    \hline     
         layout with texts but without any other class elements  &text &$\left\{underlay, logo, embellishment \right\}$\\
    \hline     
         layout with underlays but without embellishments and logos  &underlay &$\left\{logo, embellishment \right\}$ \\
    \hline
         layout with logos but without embellishments    &logo &$\left\{embellishment \right\}$\\
    \hline
         layout with embellishments &embellishment &$\left\{ \right\}$\\
    \hline
    \end{tabular}
    }
    }
    \label{tab:attribute discription}
\end{table*}

\section{Related Work}
% Graphic layout designs play a crucial role as the foundation for multimedia, such as posters ~\cite{DBLP:journals/tvcg/ODonovanAH14,DBLP:conf/ijcai/ZhouXMGJX22,DBLP:conf/cvpr/XuZGJX23,DBLP:journals/corr/abs-2303-15937} and magazines ~\cite{DBLP:journals/tomccap/YangMXRL16,DBLP:conf/siggraph/TabataYMY19,DBLP:journals/tog/ZhengQCL19,DBLP:conf/eccv/LeeJELG0Y20}. Recently, a lot of research efforts have been devoted to the graphic layout generation task. These approaches can be classified into two types: image-agnostic and image-aware layout generation, depending on whether they consider the image content.

Continuous research efforts~\cite{DBLP:journals/tvcg/ODonovanAH14,xie2021canvasemb,DBLP:conf/ijcai/ZhouXMGJX22,yu2022layoutdetr,DBLP:conf/cvpr/XuZGJX23,xuan2023cvae,DBLP:journals/corr/abs-2303-15937} have been devoted to the graphic layout generation,  which can be divided into two categories based on their consideration of image content: image-agnostic and image-aware layout generation.

\noindent\textbf{Image-agnostic layout generation.} 
Early works~\cite{DBLP:journals/tog/JacobsLSBS03,DBLP:conf/chi/KumarTAK11,DBLP:journals/tog/CaoCL12,DBLP:journals/tvcg/ODonovanAH14,DBLP:conf/chi/ODonovanAH15} mainly utilize templates or heuristic rules to design graphic layouts and often fail to produce flexible and various layouts. Recently, an increasing number of deep-learning-based models have been developed for generating graphic layouts~\cite{DBLP:conf/eccv/LeeJELG0Y20,DBLP:journals/pami/LiYHZX21,DBLP:conf/iccv/GuptaLA0MS21,DBLP:conf/cvpr/YangFYW21,DBLP:journals/corr/abs-2110-06794,DBLP:conf/mm/KikuchiSOY21,DBLP:journals/tvcg/LiY0LWX21,DBLP:conf/aaai/JiangSZL022,DBLP:journals/corr/abs-2303-11589}. LayoutGAN~\cite{DBLP:journals/pami/LiYHZX21}, LayoutVAE~\cite{DBLP:conf/iccv/JyothiDHSM19}, and LayoutVTN~\cite{DBLP:conf/cvpr/ArroyoPT21} generate layouts from noise without any conditions. To meet the diverse user demands in real-world applications, several conditional methods~\cite{kong2022blt,DBLP:journals/corr/abs-2301-11529,jiang2023layoutformer++,inoue2023layoutdm,levi2023dlt,fan2023real} have been proposed to guide the layout generation process. The condition includes graphic layout element types, numbers, sizes, and locations. For example, AttributeGAN~\cite{DBLP:journals/tvcg/LiY0LWX21} incorporates elements' aspect ratio and location as conditions to generate graphic layouts. LayoutFormer++~\cite{jiang2023layoutformer++} utilizes sequence-based control mechanisms to facilitate flexible and varied layout generation. However, the aforementioned methods primarily concentrate on modeling the internal relationships among graphic layout elements, while neglecting the connection between the graphic layout and the image content.

\noindent\textbf{Image-aware layout generation.} 
ContentGAN~\cite{DBLP:journals/tog/ZhengQCL19} combines visual information to generate layouts for magazine pages, but it cannot fully capture the image content as global pooling is applied to feature maps. To comprehend the visual-texture content of the image, CGL-GAN~\cite{DBLP:conf/ijcai/ZhouXMGJX22} and PDA-GAN~\cite{DBLP:conf/cvpr/XuZGJX23} combine CNN and transformer~\cite{DBLP:conf/nips/VaswaniSPUJGKP17} to synthesize image-aware graphic layouts for posters, which are the most relevant works in this discipline. Recently, more image-aware layout generation methods have been proposed for modeling the relationship between graphic layouts and image contents~\cite{DBLP:conf/icmcs/ZhangLW20,DBLP:journals/tmm/LiZW22,DBLP:conf/mm/CaoMZLXGJ22,DBLP:journals/corr/abs-2303-15937}. However, none of these methods devote to study image-aware layout generation with layout attributes and partial layout constraints. Although CGL-GAN and PDA-GAN mentioned that they could generate layouts according to partial layout constraints to some extent, conducting numerous qualitative evaluations and observations demonstrate that generated layouts may not always conform to user constraints. Additionally, this partial layout is limited to requiring complete element information, including both class and coordinate. More importantly, these models are unable to express layout attributes.

In real-world applications, it is typically necessary to design multiple layouts based on one product image to meet various user demands for the effective presentation of advertising posters. This paper focuses on leveraging layout attributes and partial layout constraints to generate image-aware graphic layouts.

\section{Our Method} \label{our_method}
As illustrated in Fig.~\ref{fig:model}, the structure of IUC-Layout backbone network follows the design of DETR~\cite{DBLP:conf/eccv/CarionMSUKZ20}, which includes a multi-scale convolutional neural network (CNN)~\cite{DBLP:conf/cvpr/HeZRS16,DBLP:conf/cvpr/LinDGHHB17}, a transformer encoder-decoder~\cite{DBLP:conf/nips/VaswaniSPUJGKP17}, and two fully connected layers. Firstly, the multi-scale CNN extracts image feature maps. Next, the sampled four-channel Gaussian noise according to the input layout attribute is connected with the feature maps at the final layer of the CNN, and sent to the transformer encoder for the embedding. For partial layout constraints, we encode the information of each element, i.e. its position and class, into a feature vector with the same dimension as the learned queries in DETR. Afterward, these vectors are added to the queries of the transformer decoder to guide the layout generation. In this manner, the partial layout constraint can be disabled by setting the element feature vectors to $0$. Our system allows the user to specify no more than $10$ elements in a partial layout. Finally, two fully connected layers respectively predict classes and bounding boxes of elements. 
% As shown in \cref{fig:model}, IUC-Layout can generate diverse graphic layouts for one product image according to various user constraints, including the layout attribute and partial layout constraints. 

\begin{table}[!t]
    \centering
    \setlength{\tabcolsep}{0.66mm}
    \renewcommand{\arraystretch}{1.5}
    \caption{Five sets of four-channel Gaussian noise with varying means based on different layout attributes.}
    {
    \scalebox{1.1}{
    \begin{tabular}{c|c}
    \hline
         \bf{layout attribute}  &\bf{means of four-channel noise}\\
    \hline     
         layout with texts \\ but without any other class elements &$(1, -1, -1, 1)$\\
    \hline     
         layout with underlays \\ but without embellishments and logos &$(1, -1, 1, -1)$ \\
    \hline
         layout with logos \\ but without embellishments &$(1, 1, -1, -1)$\\
    \hline
         layout with embellishments &$(1, 1, 1, 1)$\\
    \hline
         unspecified layout attribute &$(0, 0, 0, 0)$\\
    \hline
    \end{tabular}
    }
    }
    \label{tab:noise discription}
\end{table}

\subsection{Layout Attribute}
%原来的写法
% We conduct statistical analysis on the CGL-dataset \cite{DBLP:conf/ijcai/ZhouXMGJX22}, which contains four categories of elements listed in descending order of occurrence probability: text, underlay, logo, and embellishment. Considering the characteristics of the dataset and multimedia presentation requirements, we defined four layout attributes for this task: (1) layout with embellishment, (2) layout with logo but without embellishment, (3) layout with underlay but without embellishment and logo, and (4) layout with text but without any other type elements. We refer to the type that is required to appear in the generated layout as the attribute element. For instance, logo is the attribute element of "layout with logo but without embellishment". These four layout attributes correspond to four sets of four-channel Gaussian noise with different means. The sampled noise is concatenated with the feature maps extracted by the multi-scale CNN and fed into the transformer encoder. More implementation details of sampling noise can be found in the supplementary materials.

According to the factors involved in the graphical design, such as layout styles, relationships between elements and user requirements, and the corresponding statistics of CGL-Dataset, we define four types of layout attribute constraints as follows: (1) layout with texts but without any other class elements, (2) layout with underlays but without embellishments and logos, (3) layout with logos but without embellishments, and (4) layout with embellishments. Each layout attribute includes the attribute element and non-desirable elements, as illustrated in Tab.~\ref{tab:attribute discription}. 

The distinctions in layout attributes arise from elements' classes, but the number and arrangement of elements can vary. Similarly to~\cite{DBLP:conf/cvpr/KarrasLA19}, it is necessary to use multidimensional noise to represent the layout attribute since this representation assigns layout attribute with a region. Moreover, sampling noise in the spatial domain can improve the model's generalization capability and fault tolerance, enhancing the robustness of the model during training. 

As indicated in Tab.~\ref{tab:noise discription}, our model pre-assigns four sets of four-channel Gaussian noise corresponding to aforementioned layout attributes, along with one set for the unspecified attribute. The mean values of these five sets of four-channel Gaussian noise are (1, -1, -1, 1), (1, -1, 1, -1), (1, 1, -1, -1), (1, 1, 1, 1), and (0, 0, 0, 0) with a variance of 1 for each channel. These points in four-dimensional space are equidistant from each other and from the origin (0, 0, 0, 0). This design effectively balances the model's learning of different layout attribute constraints.
% Further implementation details of noise sampling and attributes analysis are available in supplementary materials.
When a user specifies a certain attribute constraint, the model samples Gaussian noise in four-dimensional space according to the corresponding noise mean and variance. The length and width of each dimension of the noise vector are equal to the input feature map size of the transformer module.

To better align with the attribute element and disentangle different attributes, we design attribute-consistent loss and attribute-disentangled loss to ensure the generated layout from the corresponding Gaussian noise satisfies the specified layout attribute constraint.

\noindent\textbf{Layout attribute-consistent loss.} We propose an attribute-consistent loss such that the generated layout contains the attribute element. Specifically, to make the attribute-consistent loss differential, we design a modified softmax function to approximately count the element number of each class $c$:
\begin{equation} \label{eq1}
    N_{c} = \sum^{Q}_{q=1}\frac{\boldsymbol{e}^{z^{c}_q \cdot \varepsilon}}{\sum_{k\in{\mathcal{K}}}\boldsymbol{e}^{z^{k}_q \cdot \varepsilon}},
\end{equation}
where $Q$ is the total number of output elements, and $z^{c}_q$ represents the output class $c$ of $q_{th}$ element. The hyperparameter $\varepsilon$ is used to increase the distinction between predicted probabilities of different classes. We set the value of $\varepsilon$ as $100$ during the training process. $\mathcal{K}$ is $\left\{text, logo, underlay, embellishment, none\right\}$. Therefore, we can calculate the layout attribute-consistent loss as:
\begin{equation} \label{eq2}
    L_{Ac} = \max{(1-N_{a}, \quad 0)},
\end{equation}
where $N_{a}$ is the number of attribute element $a$, computed by the Eq.~\ref{eq1}. If the number of attribute elements exceeds 1, $L_{Ac}$ is 0. Otherwise, $L_{Ac}$ equals ($1-N_{a}$).

\noindent\textbf{Layout attribute-disentangled loss.} If only introducing $L_{Ac}$, generated layouts may contain undesired elements. To satisfy different attribute constraints, we design attribute-disentangled losses to separate the relationships between various classes. They can be formulated uniformly as:
\begin{equation} \label{eq3}
    L^{*}_{Ad} = \sum_{u\in{\mathcal{S}^*}}N_{u},
\end{equation}
where $u$ represents the undesired element. $\mathcal{S}^*$, as indicated in Tab.~\ref{tab:attribute discription} represents the set of classes that are absent in generated layouts based on the specified attribute $*$. 
% $\mathcal{S}^*$ represents the set of categories that are required to not occur in the generated layout according to the specified attribute $*$. For instance, when the specified attribute is "layout with text but without any other type elements", $\mathcal{S}^t$ includes categories of underlay, logo, and embellishment, abbreviated as $\left\{und, log, emb\right\}$. The attribute-disentangled loss $L^t_{Ad}$ can be computed as:
% \begin{equation} \label{eq4}
%     L^t_{Ad} = N_{und} + N_{log} + N_{emb}
% \end{equation}
% Note that the specified attribute of "layout with embellishment" allows the generated layout to contain any category, $Set^e$ is a null set, thereby the attribute disentangled loss $L^e_{Ad}$ is always 0. During the training process, the corresponding $L^*_{Ad}$ is selected according to the specified attribute $*$.

\subsection{Partial Layout}
The partial layout constraints can be divided into two categories: one consists of elements with complete information, and the other contains elements with incomplete information. The reason to integrate elements of incomplete information is to enable the flexibility and diversity of partial layout constraints. For example, users may provide the elements with complete information or position only, or the mix of elements with complete information and elements of position information, etc. In this section, we introduce the partial-constraint loss $L_{P}$ and a random mask $PL_{rm}$ operation to obtain training examples of elements with incomplete information. 
% After conducting extensive evaluations and observations, we discovered that the layouts generated by CGL-GAN and PDA-GAN have deviations from partial layout, despite some predicted elements being similar to given constraints to a certain extent.

The $L_{P}$ is used to train the network to produce layouts consistent with the input partial layout constraints. It is formulated as:
\begin{equation} \label{eq5}
    L_{P} = \left\vert Pred \cdot PL_{bm} - PL  \right\vert
\end{equation}
where $Pred$ denotes the output of the layout generation model, and  $PL_{bm}$ is the binary matrix derived from the input partial layout $PL$. When a value in $PL$ is non-zero, the corresponding entry in $PL_{bm}$ is set to $1$; otherwise, it is set to $0$. The L1 distance is employed to calculate the value of $L_{P}$. The element correspondence between $pred$ and $PL$ is defined by the query index. Specifically, since we add the feature vector of the first element in a partial layout to the first query, that element should then correspond to the element produced by the first query. The rest element correspondences are done in the same way.

To augment the training with incomplete element information, we generate a random mask $PL_{rm}$ with the same size as the partial layout, consisting of $0$ and $1$. The percentage of value $0$ amounts to 25\%. $PL_{rm}$ randomly masked the information of the element class and coordinates in the partial layout. Similar to Eq.~\ref{eq6}, we can calculate the loss of partial layout constraints with the random mask as follows:
\begin{equation} \label{eq6}
    L_{PL_{rm}} = \left\vert Pred \cdot PL_{bm} \cdot PL_{rm} - PL \cdot PL_{rm}  \right\vert
\end{equation}

The proposed partial-constraint loss and random mask are simple and can be easily applied to other models. In the experimental section, we will demonstrate their effectiveness by incorporating them into CGL-GAN and PDA-GAN.

 \begin{table*}[t!]
    \centering
    \setlength{\tabcolsep}{2.2mm}
    \renewcommand{\arraystretch}{1.5}
    \caption{{\bf{Quantitative evaluation for content-aware methods.}} Bold numbers denote the best result. $\downarrow$ (or $\uparrow$) means the smaller (or bigger) value, the better. $None$ means the model lacks attribute control ability. $Text$ as a sample refers to the attribute "layout with texts but without any other class elements". $Unspecified$ means unspecified layout attributes. $\times$ represents that the model cannot complete the corresponding task, while $-$ indicates that the model does not need to be tested for the metric.}
    {
    \scalebox{1.1}{
    \begin{tabular}{ll|c|ccc|cccc}
    \hline
         Model &$Attribute$  &$R_{lac}\uparrow$  &$R_{com}\downarrow$ &$R_{shm}\downarrow$ &$R_{sub}\downarrow$ &$R_{ove}\downarrow$ &$R_{und}\uparrow$ &$R_{ali}\downarrow$ &$R_{occ}\uparrow$\\
    \hline
         ContentGAN~\cite{DBLP:journals/tog/ZhengQCL19} &$None$ &$\times$     &45.59 &17.08 &1.143 &0.0397 &0.8626 &0.0071 &93.4\\
         CGL-GAN~\cite{DBLP:conf/ijcai/ZhouXMGJX22} &$None$    &$\times$     &35.77 &15.47 &0.805 &0.0233 &0.9359 &0.0098 &99.6\\
         PDA-GAN~\cite{DBLP:conf/cvpr/XuZGJX23} &$None$   &$\times$      &33.55 &12.77 &0.688 &0.0290 &\bf{0.9481} &0.0105 &99.7\\
    \hline
         IUC-Layout (Ours) &$Text$  &0.973 &34.23 &\bf{10.43} &\bf{0.664} &\bf{0.0129} &$-$ &0.0084  &97.2\\
         IUC-Layout (Ours) &$Underlay$ &0.980 &\bf{32.69} &16.64 &0.816 &0.0172 &0.9312 &\bf{0.0030}  &99.8\\
         IUC-Layout (Ours) &$Logo$ &1.000 &35.93 &16.27 &0.936 &0.0255 &0.9226 &0.0144  &\bf{100.0}\\
         IUC-Layout (Ours) &$Embellishment$ &0.996 &33.79 &15.66 &0.899 &0.0291 &0.9163 &0.0081 &\bf{100.0}\\
         IUC-Layout (Ours) &$Unspecified$ &$-$   &33.06 &15.93 &0.826 &0.0174 &0.9221 &0.0055 &99.9\\
    \hline
    \end{tabular}
    }
    }
    \label{tab:content-aware}
\end{table*}

\begin{figure*}[!t]
    \centering
    \hspace{0cm}\includegraphics[width=\textwidth]{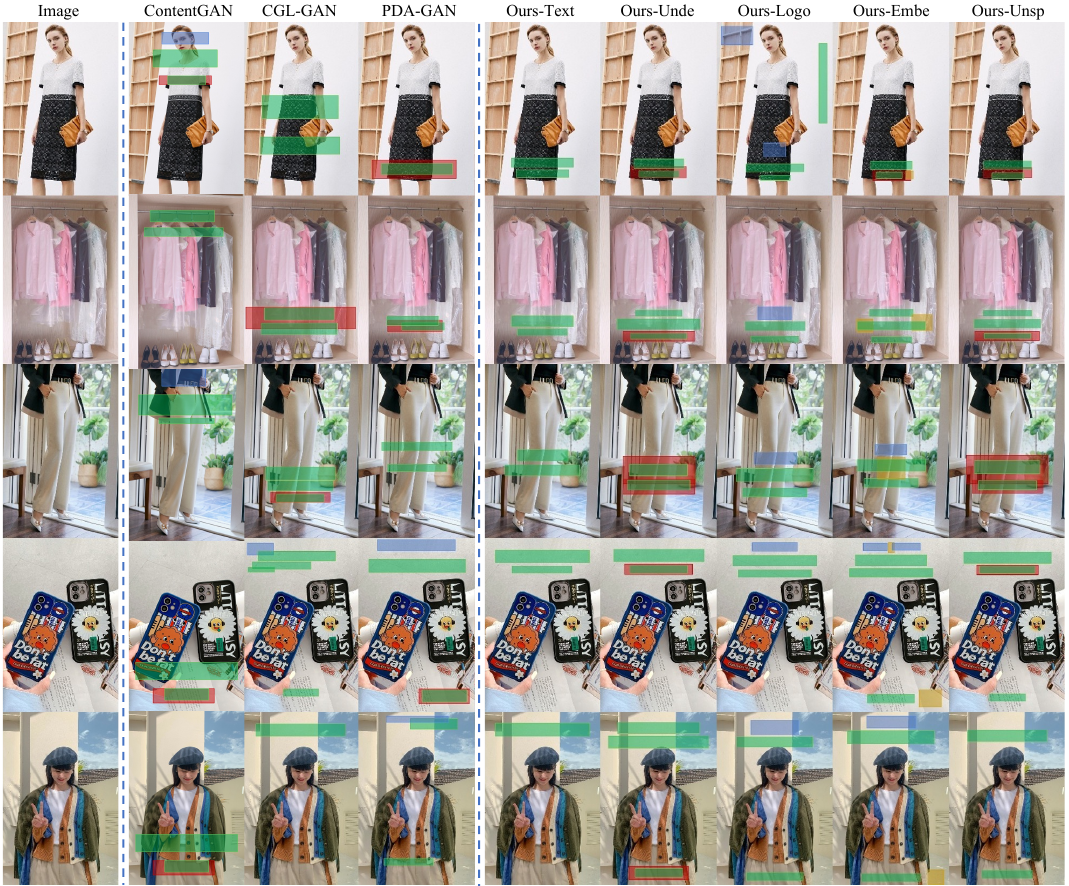}
    \caption{{\bf Qualitative evaluation for image-aware models.} The layouts in each row are conditioned on the same product image, while the ones in a column are generated by the same model. $Ours\mbox{-}Unsp$ represents unspecified attributes in our model.}
    \label{fig:ep1}
\vspace{-0mm}
\end{figure*}

\section{Experiments}
This section primarily compares our model with SOTA layout generation methods and presents its ablation studies. Given space limitations, more experimental comparisons, including a user study and analyses of computational complexity for different models, can be found in the supplementary material.

\subsection{Implementation Details}
We implement our model in PyTorch 1.7.1 and utilize the Adam optimizer~\cite{DBLP:journals/corr/KingmaB14} for training. Initial learning rates are set to $10^{-5}$ for CNN, and $10^{-4}$ for the transformer and fully connected layers. We take CGL-Dataset as both training and test datasets. To ensure fair comparisons, following CGL-GAN and PDA-GAN, we resize the input image to $240 \times 350$. 
% Compared to CGL-GAN and PDA-GAN, our model has the lowest number of parameters ($3.773\times10^7$) and FLOPs ($1.528\times10^{11}$). 
Our model is trained for 300 epochs with a batch size of 128. Learning rates are reduced by a factor of 10 after 200 epochs. The total training time is approximately 37.8 hours, utilizing 16 NVIDIA V100 GPUs.

During training, we combine four loss functions: $L_{rec}$, $L_{Ac}$, $L^*_{Ad}$, and $L_{PL_{rm}}$, to guide the model optimization. $R_{rec}$ is the reconstruction loss that penalizes the deviation between the ground truth and the generated layout. We calculate $L_{rec}$ following~\cite{DBLP:conf/eccv/CarionMSUKZ20}. The overall training loss for the model can be summed as follows:
\begin{equation} \label{eq7}
    L = L_{rec} + \beta \cdot L_{Ac} + \gamma \cdot L^*_{Ad} + \eta \cdot L_{PL_{rm}}
\end{equation}
Where $\beta$, $\gamma$, and $\eta$ are three weight coefficients. By observation, we found that $L^*_{Ad}$ is approximately ten times larger than $L_{Ac}$, thus we set $\gamma$ to 0.1, both $\beta$ and $\eta$ to 1.0.

\subsection{Metrics}
For quantitative evaluations, we follow~\cite{DBLP:conf/ijcai/ZhouXMGJX22,DBLP:conf/cvpr/XuZGJX23} to adopt composition-relevant and graphic metrics to evaluate the performance of our model. Composition-relevant metrics include measuring text background complexity $R_{com}$, occlusion subject degree $R_{shm}$, and occlusion product degree $R_{sub}$. Graphic metrics consist of layout overlap $R_{ove}$, underlay overlap $R_{und}$, layout alignment $R_{ali}$, and the ratio of nonempty layouts $R_{occ}$. In addition, we introduce two metrics, $R_{lac}$ and $R_{plc}$, to evaluate the model's performance on layout attribute and partial layout constraints, respectively. $R_{lac}$ indicates the ratio of generated layouts that comply with the given attribute constraints. $R_{plc}$ is used to quantify the average difference between given partial layout constraints and generated layouts.
% and it can be formulated as:
% \begin{equation}
%     R_{plc} = \frac{1}{N} \sum^{N}_{i=1}
%     \left\vert
%     Pred_{i-id} - PL_{i-id}
%     \right\vert
% \end{equation}
% $N$ represents the total number of given partial layout information. $Pred_{i\mbox{-}id}$ represents the predicted value at the $i\mbox{-}th$ index $id$ of model output according to given partial layout index $id$. $PL_{i\mbox{-}id}$ is the value of given partial layout at the $i\mbox{-}th$ index. 
Combining the above metrics can reflect the model's performances regarding graphic quality, product content relevance, layout attribute constraint, and partial layout consistency.

\begin{table*}[!t]
    \centering
    \setlength{\tabcolsep}{2.66mm}
    \renewcommand{\arraystretch}{1.5}
    \caption{{\bf Quantitative evaluation for image-agnostic methods.}}
    {
    \scalebox{1.1}{
    \begin{tabular}{ll|c|ccc|ccc}
    \hline
         Model &$Attribute$  &$R_{lac}\uparrow$  &$R_{com}\downarrow$ &$R_{shm}\downarrow$ &$R_{sub}\downarrow$ &$R_{ove}\downarrow$ &$R_{und}\uparrow$ &$R_{ali}\downarrow$\\
    \hline
         LayoutTransformer~\cite{DBLP:conf/iccv/GuptaLA0MS21} &$None$    &$\times$   &40.92 &21.08 &1.310 &0.0156 &0.9516 &0.0049\\
         LayoutVTN~\cite{DBLP:conf/cvpr/ArroyoPT21} &$None$  &$\times$   &41.77 &22.21 &1.323 &0.0130 &\bf{0.9698} &0.0047\\
    \hline
         IUC-Layout (Ours) &$Text$ &0.973 &34.23 &\bf{10.43} &\bf{0.664} &\bf{0.0129} &$-$ &0.0084 \\
         IUC-Layout (Ours) &$Underlay$ &0.980 &\bf{32.69} &16.64 &0.816 &0.0172 &0.9312 &\bf{0.0030}  \\
         IUC-Layout (Ours) &$Logo$ &1.000 &35.93 &16.27 &0.936 &0.0255 &0.9226 &0.0144  \\
         IUC-Layout (Ours) &$Embellishment$ &0.996 &33.79 &15.66 &0.899 &0.0291 &0.9163 &0.0081   \\
         IUC-Layout (Ours) &$Unspecified$ &$-$ &33.06 &15.93 &0.826 &0.0174 &0.9221 &0.0055   \\
    \hline
    \end{tabular}
    }
    }
    \label{tab:image-agnostic}
\end{table*}

\begin{figure*}[!t]
    \centering
    \hspace{0cm}\includegraphics[width=\textwidth]{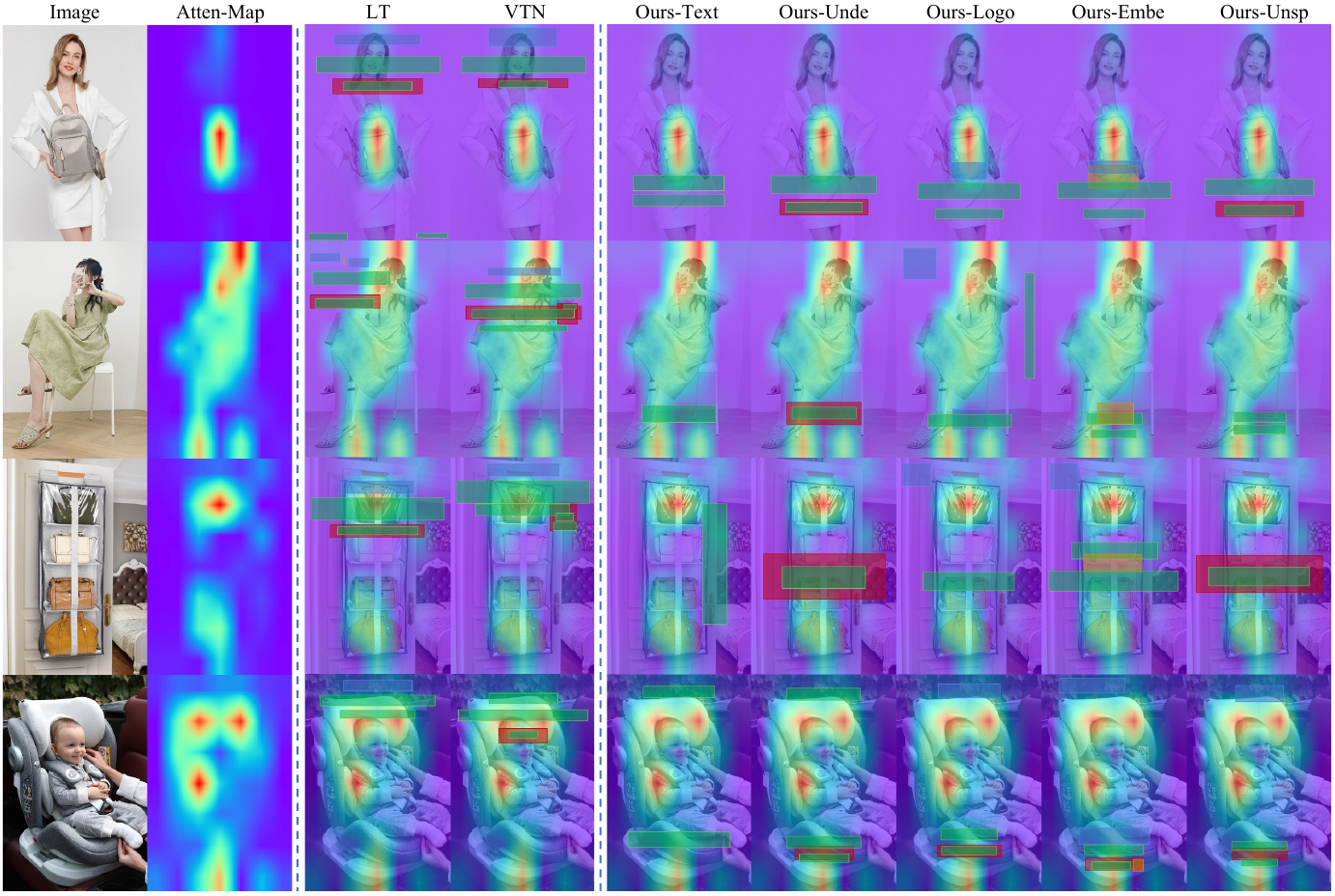}
    \caption{{\bf Qualitative evaluation for image-agnostic models.} Layouts in each row are conditioned on the same image with product attention map $Atten\mbox{-}Map$ \protect\cite{DBLP:conf/iccv/CheferGW21,DBLP:conf/icml/RadfordKHRGASAM21}. $LT$ and $VTN$ represent LayoutTransformer and LayoutVTN, respectively.}
    \label{fig:ep2-image-agnostic}
\end{figure*}

\subsection{Layout Attribute Constraints}
    \noindent\textbf{Comparison with image-aware methods.} We first compare our method with image-aware methods: ContentGAN~\cite{DBLP:journals/tog/ZhengQCL19}, CGL-GAN~\cite{DBLP:conf/ijcai/ZhouXMGJX22}, and PDA-GAN~\cite{DBLP:conf/cvpr/XuZGJX23}. Note that when the attribute is "layout with text but without any other class elements", IUC-Layout network does not generate underlays, rendering $R_{und}$ irrelevant. Quantitative evaluations presented in Tab.~\ref{tab:content-aware} demonstrate that IUC-Layout network achieves the best performance in all other metrics except for $R_{und}$. A key strength of IUC-Layout network lies in its ability to generate image-aware layouts adhering to various attribute constraints. 

Correspondingly, qualitative comparisons are presented in Fig.~\ref{fig:ep1}. Note that embellishments can overlap with any elements as their purpose is to enhance the layouts' aesthetics. Underlays are commonly used alongside texts to emphasize them. Columns 5 to 8 of Fig.~\ref{fig:ep1} demonstrate that our model effectively aligns with and disentangles distinct layout attributes. For instance, in the 7th column, all generated layouts by IUC-Layout network include the logo (consistent), while excluding the embellishment (disentangled). 

Interestingly, as shown in the 2nd, 3rd, and 4th columns in Fig.~\ref{fig:ep1}, previous works tend to create layouts with few or even no embellishment due to the low frequency (3.23\%) of the embellishments in the CGL-Dataset. In contrast, our model consistently generates embellishments when the designated attribute is "layout with embellishment", as demonstrated in the 8th column. Moreover, the 4th and 5th rows in Fig.~\ref{fig:ep1} highlight our model's superiority over CGL-GAN and PDA-GAN in terms of layout overlap and alignment. 

\noindent\textbf{Comparison with image-agnostic methods.}
We also compare IUC-Layout network with image-agnostic methods, including LayoutTransformer~\cite{DBLP:conf/iccv/GuptaLA0MS21} and LayoutVTN~\cite{DBLP:conf/cvpr/ArroyoPT21}. Quantitative results in Tab.~\ref{tab:image-agnostic} demonstrate that our model outperforms LayoutTransformer and LayoutVTN in composition-relevant metrics ($R_{com}$, $R_{shm}$, and $R_{sub}$) across all attribute conditions. As shown in Fig.~\ref{fig:ep2-image-agnostic}, layouts generated by our model are more effective in avoiding high-product-attention regions and human faces, enabling a comprehensive and visually pleasing presentation of product information. 
% Furthermore, as seen in the 1st and 2nd rows of \cref{fig:ep2-image-agnostic}, layout bounding boxes generated by LayoutTransformer and LayoutVTN may obscure sensitive areas of the image, such as the human head and face. Using such layouts to create posters would lead to poor visual effects. In contrast, layouts generated by IUC-Layout can precisely avoid the sensitive areas of the product image, improving the visual aesthetics of the resulting posters.

\begin{table*}[!t]
    \centering
    \setlength{\tabcolsep}{2.66mm}
    \renewcommand{\arraystretch}{1.5}
    \caption{{\bf Ablation studies on attribute losses.} $\checkmark$ ($\times$) denotes our model trained with (without) attribute losses $L_{A}$. Each pair of rows presents an experimental comparison under the same attribute constraint. $Text$ as a sample indicates "layout with texts but without any other class elements".} 
    {
    \scalebox{1.1}{
    \begin{tabular}{ll|c|ccc|ccc}
    \hline
         Attribute &$L_{A}$  &$R_{lac}\uparrow$  &$R_{com}\downarrow$ &$R_{shm}\downarrow$ &$R_{sub}\downarrow$ &$R_{ove}\downarrow$ &$R_{und}\uparrow$ &$R_{ali}\downarrow$\\
    \hline
         Text &$\times$     &\bf{0.981} &35.08 &11.68 &0.724 &0.0162 &$-$ &0.0122  \\
         Text &$\checkmark$  &0.973 &\bf{34.23} &\bf{10.43} &\bf{0.664} &\bf{0.0129} &$-$ &\bf{0.0084} \\
    \hline
         Underlay &$\times$     &0.958 &34.31 &17.15 &0.916 &\bf{0.0138} &0.8969 &0.0048   \\
         Underlay &$\checkmark$  &\bf{0.980} &\bf{32.69} &\bf{16.64} &\bf{0.816} &0.0172 &\bf{0.9312} &\bf{0.0030}  \\
    \hline
         Logo   &$\times$     &0.998 &36.68 &17.09 &0.975 &0.0529 &0.9038 &0.0157    \\
         Logo   &$\checkmark$  &\bf{1.000} &\bf{35.93} &\bf{16.27} &\bf{0.936} &\bf{0.0255} &\bf{0.9226} &\bf{0.0144}  \\
    \hline
         Embellishment &$\times$     &0.989 &35.25 &\bf{15.01} &0.916 &\bf{0.0261} &0.9035 &\bf{0.0079}    \\
         Embellishment &$\checkmark$  &\bf{0.996} &\bf{33.79} &15.66 &\bf{0.899} &0.0291 &\bf{0.9163} &0.0081   \\
    \hline
         Unspecified &$\times$     &$-$ &35.01 &\bf{15.19} &0.883 &0.0321 &0.8914 &0.0092  \\
         Unspecified &$\checkmark$  &$-$ &\bf{33.06} &15.93 &\bf{0.826} &\bf{0.0174} &\bf{0.9221} &\bf{0.0055}   \\
    \hline
    \end{tabular}
    }
    }
    \label{tab:ablation-AL}
\end{table*}

\begin{table}[!t]
    \centering
    \setlength{\tabcolsep}{2.6mm}
    \renewcommand{\arraystretch}{1.5}
    \caption{{\bf Quantitative evaluation on $L_{P}$ and random mask $PL_{rm}$.} $IcEI$ means whether the method can handle partial layout constraints with \emph{incomplete element information}.}
    {
    \scalebox{1.1}{
    \begin{tabular}{lcc|cc}
    \hline
         Model  &$L_{P}$  &$PL_{rm}$  &$R_{plc}\downarrow$ &$IcEI$ \\
    \hline
         CGL-GAN~\cite{DBLP:conf/ijcai/ZhouXMGJX22} &\quad &\quad    &0.2895  &$\times$\\
         PDA-GAN~\cite{DBLP:conf/cvpr/XuZGJX23} &\quad &\quad    &0.2693  &$\times$\\
    \hline
         CGL-GAN~\cite{DBLP:conf/ijcai/ZhouXMGJX22} &\checkmark  &\quad   &0.0004  &$\times$\\
         PDA-GAN~\cite{DBLP:conf/cvpr/XuZGJX23} &\checkmark  &\quad   &0.0004  &$\times$\\
    \hline
         CGL-GAN~\cite{DBLP:conf/ijcai/ZhouXMGJX22} &\checkmark &\checkmark    &0.0006  &\checkmark\\
         PDA-GAN~\cite{DBLP:conf/cvpr/XuZGJX23} &\checkmark &\checkmark    &0.0008  &\checkmark\\
    \hline
    \end{tabular}
    }
    }
    \label{tab:user constraints}
\end{table}

\begin{figure*}[!t]
\centering
\includegraphics[width=\textwidth]{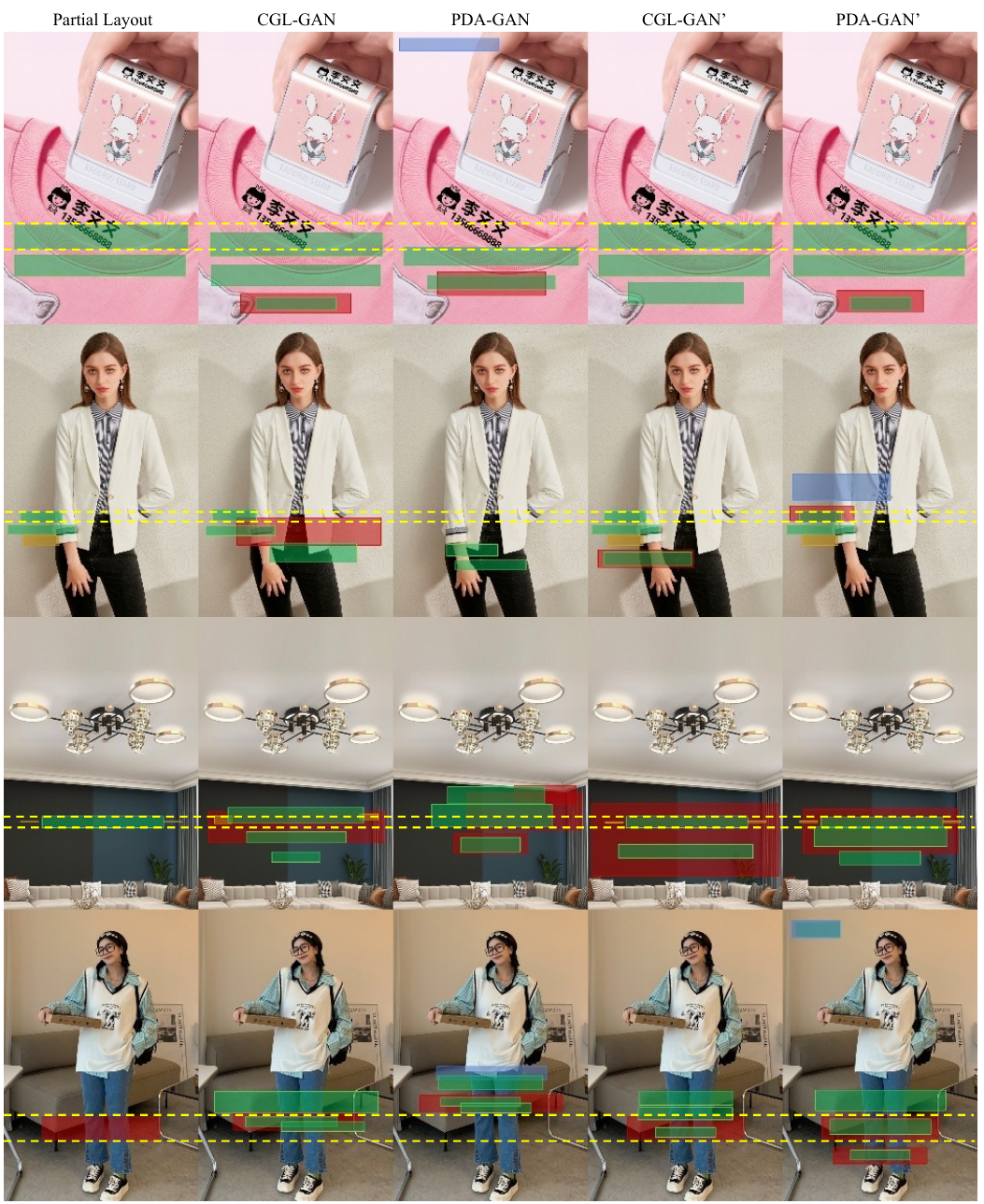}
\caption{{\bf Effects of $L_{P}$.} The yellow dashed line is used to measure the alignment between the generated layouts and the given partial layout. $CGL\mbox{-}GAN'$ and $PDA\mbox{-}GAN'$ mean CGL-GAN and PDA-GAN with $L_{P}$, respectively.}
\label{fig:exp_constraints}
\vspace{10mm}
\end{figure*}

\begin{figure*}[!t]
\centering
\includegraphics[width=\textwidth]{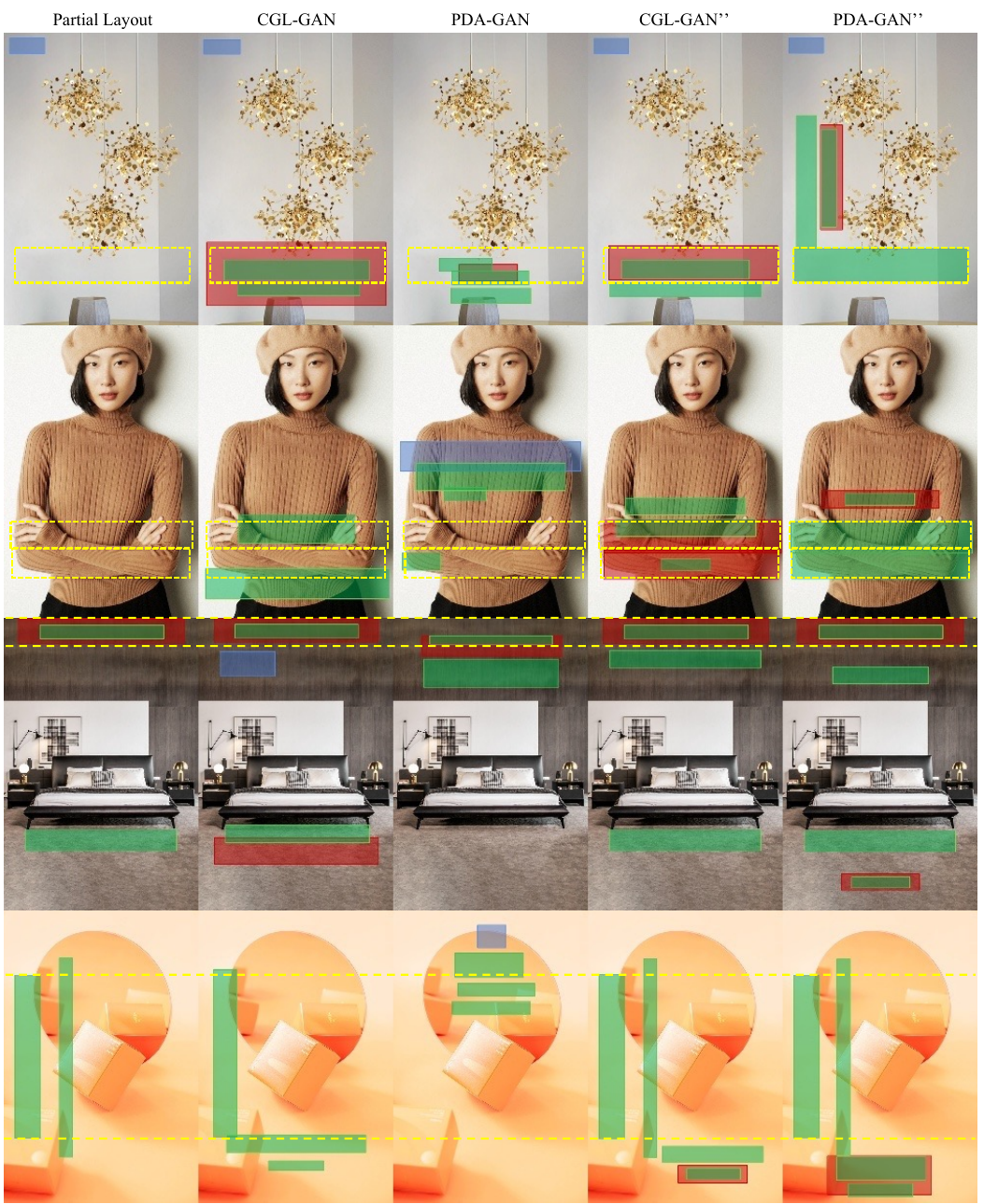}
\caption{{\bf Effects of $L_{PL_{rm}}$.} The yellow boxes in the first two rows indicate the element with box coordinates but without class information. $CGL\mbox{-}GAN''$ and $PDA\mbox{-}GAN''$ mean CGL-GAN and PDA-GAN with $L_{PL_{rm}}$, respectively.}
\label{fig:exp_constraints_RM}
\vspace{10mm}
\end{figure*}

\noindent\textbf{Ablation studies for attribute losses.} 
To validate the effectiveness of designed attribute losses, we conducted paired ablation studies for each attribute constraint. As shown in Tab.~\ref{tab:ablation-AL}, after incorporating attribute losses, the model exhibited a clear advantage in both composition-relevant and graphic metrics. 
% These five sets of comparative experiments provide strong evidence for the substantial benefits brought to the model by the attribute losses.

Additionally, we explored multiple ablation experiments on the attribute constraint module. One approach involved utilizing a linear layer to process a single value and connect it with feature maps, while another method replaced the Gaussian noise with the fixed mean. Unfortunately, neither of these methods yielded the anticipated results. Due to space constraints, we provide more comparative sample presentations in the supplementary material.

\subsection{Partial layout Constraints}
\noindent\textbf{Effects of $L_{P}$.}
As shown in the first four rows of Tab.~\ref{tab:user constraints}, introducing $L_{P}$ to CGL-GAN (PDA-GAN) reduces the dissimilarity between generated layouts and provided information from 0.2895 (0.2693) to 0.0004 (0.0004). Layouts generated by CGL-GAN and PDA-GAN in the 1st and 3rd rows of Fig.~\ref{fig:exp_constraints} roughly match the given element's information but with distinct positional deviations. In contrast, models with $L_{P}$ produce layouts that closely adhere to provided constraints. In the 2nd row, models without $L_{P}$ generate layouts that even missing some elements from the partial layout. Notably, from the 4th and 5th columns, models with $L_{P}$ generate layouts that not only maintain high consistency with the given partial layout but also precisely align newly generated element boxes with the provided element positions, effectively enhancing the layout aesthetics. Furthermore, in the 3rd and 4th rows, when the partial layout includes underlay (or text) element, the model can correspondingly generate text (or underlay), resulting in a harmonized layout.

\noindent\textbf{Effects of $PL_{rm}$.} 
As shown in Tab.~\ref{tab:user constraints}, previous models without $PL_{rm}$ cannot handle partial layout constraints with incomplete element information. The introduction of $PL_{rm}$ slightly reduces performance compared to using $L_{P}$ only, possibly due to the increased complexity of the task by $PL_{rm}$. In Fig.~\ref{fig:exp_constraints_RM}, the first partial layout includes a logo element and coordinates of another element without class information. The second sample provides the positions of two boxes without classes. The first two samples in Fig.~\ref{fig:exp_constraints_RM} demonstrate that layouts generated by models with $L_{PL_{rm}}$ are consistent with complete elements and reasonably supplement incomplete elements. The last two rows show that models with $L_{PL_{rm}}$ also perform well in partial layout constraints with complete element information.

% More comparative samples to show the effectiveness of $L_{P}$ and $PL_{rm}$ can be found in supplementary materials.

\begin{table}[t!]
    \vspace{0pt}
    \vspace{0pt}
    \centering
    \setlength{\tabcolsep}{2.88mm}
    \renewcommand{\arraystretch}{1.5}
    \caption{Cost comparison.}
    {
    \scalebox{1.1}{
    \begin{tabular}{l|c|c}
    \hline
         Model  &Parameters  &FLOPs \\
    \hline     
         CGL-GAN~\cite{DBLP:conf/ijcai/ZhouXMGJX22} &$5.745\times10^7$ &$2.274\times10^{11}$\\
         PDA-GAN~\cite{DBLP:conf/cvpr/XuZGJX23} &$3.806\times10^7$ &$1.929\times10^{11}$\\
         IUC-Layout (Ours) &$\bf{3.773}\times\bf{10^7}$ &$\bf{1.528}\times\bf{10^{11}}$\\
    \hline
    \end{tabular}
    }
    \vspace{0pt}
    }
    \label{tab:cost}
\end{table}

\begin{table}[t!]
    \vspace{0pt}
    \vspace{0pt}
    \centering
    \setlength{\tabcolsep}{2.88mm}
    \renewcommand{\arraystretch}{1.5}
    \caption{User study. * denotes the professional group.}
    {
    \scalebox{1.1}{
    \begin{tabular}{l|cccc}
    \hline
         Model  &$P_{e}\uparrow$ &$P_{b}\uparrow$  &$P^{*}_{e}\uparrow$ &$P^{*}_{b}\uparrow$\\ 
    \hline
         CGL-GAN~\cite{DBLP:conf/ijcai/ZhouXMGJX22} &26.96 &20.83 &26.39 &22.74\\
         PDA-GAN~\cite{DBLP:conf/cvpr/XuZGJX23} &26.55 &23.13  &26.03 &21.07\\
         IUC-Layout (Ours) &\bf{46.49} &\bf{56.04} &\bf{47.58} &\bf{56.19}\\
    \hline
    \end{tabular}}}
    \vspace{0pt}
    \label{tab:user study}
\end{table}

\subsection{Computational complexity and user study.}
\noindent\textbf{Computational complexity.} 
As shown in Tab.~\ref{tab:cost}, compared to other image-aware layout generation models, our model has the lowest number of parameters ($3.773\times10^7$) and computational complexity ($1.528\times10^{11}$). In testing, it only needs 4.1 ms to yield a layout on one NVIDIA V100 GPU. This signifies the suitability of our model for practical implementation.

\noindent\textbf{User study.} 
In addition to the general quantitative metrics, we also conducted a user study, as shown in Tab.~\ref{tab:user study}, to accurately evaluate the model's performance. We randomly selected 60 test samples (20 with no user constraints, 20 with attribute constraints, and 20 with partial layout constraints). Each sample includes one product image and three corresponding predicted layouts (by CGL-GAN, PDA-GAN, and our model). We split participants into two groups (5 professional designers and 24 novice designers) and asked them to select eligible and best layouts from the three predicted layouts. The eligible-selected (best-selected) layout percentage $P_e$ ($P_b$), which is the ratio of this model's vote count to the total vote count of all models, are shown in Tab.~\ref{tab:user study}, revealing that our model's performance significantly outperformed other methods.

% In addition to the aforementioned evaluations, we conducted a user study in the supplementary materials, demonstrating that our model received superior assessments. Further experimental details can be found in the supplementary.
In addition to the aforementioned evaluations, further experimental details can be found in the supplementary.

\section{Conclusion}
Our proposed IUC-Layout network is designed for generating image-aware layouts with diverse user constraints. To generate layouts with specified attributes, we propose attribute-consistent and attribute-disentangled losses. We also introduce a random mask and partial-layout loss to satisfy partial layout constraints, which can be easily applied to other methods. Quantitative and qualitative evaluations demonstrate that IUC-Layout network can generate high-quality image-aware layouts that adhere to various user constraints. In the future, we will further explore image-aware layout generation with intuitive user constraints, including but not limited to sequence constraints.

% \section*{Acknowledgments}
% This should be a simple paragraph before the References to thank those individuals and institutions who have supported your work on this article.

\bibliographystyle{IEEEtran}
\bibliography{cite}

@article{DBLP:journals/pami/LiYHZX21,
  author       = {Jianan Li and
                  Jimei Yang and
                  Aaron Hertzmann and
                  Jianming Zhang and
                  Tingfa Xu},
  title        = {LayoutGAN: Synthesizing Graphic Layouts With Vector-Wireframe Adversarial
                  Networks},
  journal      = {{IEEE} Trans. Pattern Anal. Mach. Intell.},
  volume       = {43},
  number       = {7},
  pages        = {2388--2399},
  year         = {2021}
}

@inproceedings{DBLP:conf/ijcai/ZhouXMGJX22,
  author       = {Min Zhou and
                  Chenchen Xu and
                  Ye Ma and
                  Tiezheng Ge and
                  Yuning Jiang and
                  Weiwei Xu},
  title        = {Composition-aware Graphic Layout {GAN} for Visual-Textual Presentation
                  Designs},
  booktitle    = {{IJCAI}},
  pages        = {4995--5001},
  publisher    = {ijcai.org},
  year         = {2022}
}

@inproceedings{DBLP:conf/iccv/JyothiDHSM19,
  author       = {Akash Abdu Jyothi and
                  Thibaut Durand and
                  Jiawei He and
                  Leonid Sigal and
                  Greg Mori},
  title        = {LayoutVAE: Stochastic Scene Layout Generation From a Label Set},
  booktitle    = {{ICCV}},
  pages        = {9894--9903},
  publisher    = {{IEEE}},
  year         = {2019}
}

@inproceedings{DBLP:conf/cvpr/ArroyoPT21,
  author       = {Diego Mart{\'{\i}}n Arroyo and
                  Janis Postels and
                  Federico Tombari},
  title        = {Variational Transformer Networks for Layout Generation},
  booktitle    = {{CVPR}},
  pages        = {13642--13652},
  publisher    = {Computer Vision Foundation / {IEEE}},
  year         = {2021}
}

@inproceedings{DBLP:conf/iccv/GuptaLA0MS21,
  author       = {Kamal Gupta and
                  Justin Lazarow and
                  Alessandro Achille and
                  Larry Davis and
                  Vijay Mahadevan and
                  Abhinav Shrivastava},
  title        = {LayoutTransformer: Layout Generation and Completion with Self-attention},
  booktitle    = {{ICCV}},
  pages        = {984--994},
  publisher    = {{IEEE}},
  year         = {2021}
}

@article{DBLP:journals/tog/ZhengQCL19,
  author       = {Xinru Zheng and
                  Xiaotian Qiao and
                  Ying Cao and
                  Rynson W. H. Lau},
  title        = {Content-aware generative modeling of graphic design layouts},
  journal      = {{ACM} Trans. Graph.},
  volume       = {38},
  number       = {4},
  pages        = {133:1--133:15},
  year         = {2019}
}

@inproceedings{DBLP:conf/mm/CaoMZLXGJ22,
  author       = {Yunning Cao and
                  Ye Ma and
                  Min Zhou and
                  Chuanbin Liu and
                  Hongtao Xie and
                  Tiezheng Ge and
                  Yuning Jiang},
  title        = {Geometry Aligned Variational Transformer for Image-conditioned Layout
                  Generation},
  booktitle    = {{ACM} Multimedia},
  pages        = {1561--1571},
  publisher    = {{ACM}},
  year         = {2022}
}

@article{DBLP:journals/corr/abs-2303-15937,
  author       = {HsiaoYuan Hsu and
                  Xiangteng He and
                  Yuxin Peng and
                  Hao Kong and
                  Qing Zhang},
  title        = {PosterLayout: {A} New Benchmark and Approach for Content-aware Visual-Textual
                  Presentation Layout},
  journal      = {CoRR},
  volume       = {abs/2303.15937},
  year         = {2023}
}

@article{DBLP:journals/tvcg/LiY0LWX21,
  author       = {Jianan Li and
                  Jimei Yang and
                  Jianming Zhang and
                  Chang Liu and
                  Christina Wang and
                  Tingfa Xu},
  title        = {Attribute-Conditioned Layout {GAN} for Automatic Graphic Design},
  journal      = {{IEEE} Trans. Vis. Comput. Graph.},
  volume       = {27},
  number       = {10},
  pages        = {4039--4048},
  year         = {2021},
  url          = {https://doi.org/10.1109/TVCG.2020.2999335},
  doi          = {10.1109/TVCG.2020.2999335},
  bibsource    = {dblp computer science bibliography, https://dblp.org}
}

@article{DBLP:journals/corr/abs-2303-05049,
  author       = {Mude Hui and
                  Zhizheng Zhang and
                  Xiaoyi Zhang and
                  Wenxuan Xie and
                  Yuwang Wang and
                  Yan Lu},
  title        = {Unifying Layout Generation with a Decoupled Diffusion Model},
  journal      = {CoRR},
  volume       = {abs/2303.05049},
  year         = {2023}
}

@inproceedings{DBLP:conf/eccv/LeeJELG0Y20,
  author       = {Hsin{-}Ying Lee and
                  Lu Jiang and
                  Irfan Essa and
                  Phuong B. Le and
                  Haifeng Gong and
                  Ming{-}Hsuan Yang and
                  Weilong Yang},
  title        = {Neural Design Network: Graphic Layout Generation with Constraints},
  booktitle    = {{ECCV} {(3)}},
  series       = {Lecture Notes in Computer Science},
  volume       = {12348},
  pages        = {491--506},
  publisher    = {Springer},
  year         = {2020}
}

@inproceedings{DBLP:conf/siggraph/TabataYMY19,
  author       = {Sou Tabata and
                  Hiroki Yoshihara and
                  Haruka Maeda and
                  Kei Yokoyama},
  title        = {Automatic layout generation for graphical design magazines},
  booktitle    = {{SIGGRAPH} Posters},
  pages        = {9:1--9:2},
  publisher    = {{ACM}},
  year         = {2019}
}

@article{DBLP:journals/tvcg/ODonovanAH14,
  author       = {Peter O'Donovan and
                  Aseem Agarwala and
                  Aaron Hertzmann},
  title        = {Learning Layouts for Single-PageGraphic Designs},
  journal      = {{IEEE} Trans. Vis. Comput. Graph.},
  volume       = {20},
  number       = {8},
  pages        = {1200--1213},
  year         = {2014}
}

@article{DBLP:journals/tog/CaoCL12,
  author       = {Ying Cao and
                  Antoni B. Chan and
                  Rynson W. H. Lau},
  title        = {Automatic stylistic manga layout},
  journal      = {{ACM} Trans. Graph.},
  volume       = {31},
  number       = {6},
  pages        = {141:1--141:10},
  year         = {2012}
}

@article{DBLP:journals/tog/JacobsLSBS03,
  author       = {Charles E. Jacobs and
                  Wilmot Li and
                  Evan Schrier and
                  David Bargeron and
                  David Salesin},
  title        = {Adaptive grid-based document layout},
  journal      = {{ACM} Trans. Graph.},
  volume       = {22},
  number       = {3},
  pages        = {838--847},
  year         = {2003}
}

@inproceedings{DBLP:conf/chi/KumarTAK11,
  author       = {Ranjitha Kumar and
                  Jerry O. Talton and
                  Salman Ahmad and
                  Scott R. Klemmer},
  title        = {Bricolage: example-based retargeting for web design},
  booktitle    = {{CHI}},
  pages        = {2197--2206},
  publisher    = {{ACM}},
  year         = {2011}
}

@inproceedings{DBLP:conf/mm/KikuchiSOY21,
  author       = {Kotaro Kikuchi and
                  Edgar Simo{-}Serra and
                  Mayu Otani and
                  Kota Yamaguchi},
  title        = {Constrained Graphic Layout Generation via Latent Optimization},
  booktitle    = {{ACM} Multimedia},
  pages        = {88--96},
  publisher    = {{ACM}},
  year         = {2021}
}

@inproceedings{DBLP:conf/cvpr/YangFYW21,
  author       = {Cheng{-}Fu Yang and
                  Wan{-}Cyuan Fan and
                  Fu{-}En Yang and
                  Yu{-}Chiang Frank Wang},
  title        = {LayoutTransformer: Scene Layout Generation With Conceptual and Spatial
                  Diversity},
  booktitle    = {{CVPR}},
  pages        = {3732--3741},
  publisher    = {Computer Vision Foundation / {IEEE}},
  year         = {2021}
}

@article{DBLP:journals/corr/abs-2303-11589,
  author       = {Junyi Zhang and
                  Jiaqi Guo and
                  Shizhao Sun and
                  Jian{-}Guang Lou and
                  Dongmei Zhang},
  title        = {LayoutDiffusion: Improving Graphic Layout Generation by Discrete Diffusion
                  Probabilistic Models},
  journal      = {CoRR},
  volume       = {abs/2303.11589},
  year         = {2023}
}

@article{DBLP:journals/corr/abs-2301-11529,
  author       = {Chin{-}Yi Cheng and
                  Forrest Huang and
                  Gang Li and
                  Yang Li},
  title        = {PLay: Parametrically Conditioned Layout Generation using Latent Diffusion},
  journal      = {CoRR},
  volume       = {abs/2301.11529},
  year         = {2023}
}

@inproceedings{DBLP:conf/eccv/CarionMSUKZ20,
  author       = {Nicolas Carion and
                  Francisco Massa and
                  Gabriel Synnaeve and
                  Nicolas Usunier and
                  Alexander Kirillov and
                  Sergey Zagoruyko},
  title        = {End-to-End Object Detection with Transformers},
  booktitle    = {{ECCV} {(1)}},
  series       = {Lecture Notes in Computer Science},
  volume       = {12346},
  pages        = {213--229},
  publisher    = {Springer},
  year         = {2020}
}

@inproceedings{DBLP:conf/cvpr/HeZRS16,
  author       = {Kaiming He and
                  Xiangyu Zhang and
                  Shaoqing Ren and
                  Jian Sun},
  title        = {Deep Residual Learning for Image Recognition},
  booktitle    = {{CVPR}},
  pages        = {770--778},
  publisher    = {{IEEE} Computer Society},
  year         = {2016}
}

@inproceedings{DBLP:conf/cvpr/LinDGHHB17,
  author       = {Tsung{-}Yi Lin and
                  Piotr Doll{\'{a}}r and
                  Ross B. Girshick and
                  Kaiming He and
                  Bharath Hariharan and
                  Serge J. Belongie},
  title        = {Feature Pyramid Networks for Object Detection},
  booktitle    = {{CVPR}},
  pages        = {936--944},
  publisher    = {{IEEE} Computer Society},
  year         = {2017}
}

@inproceedings{DBLP:conf/nips/VaswaniSPUJGKP17,
  author       = {Ashish Vaswani and
                  Noam Shazeer and
                  Niki Parmar and
                  Jakob Uszkoreit and
                  Llion Jones and
                  Aidan N. Gomez and
                  Lukasz Kaiser and
                  Illia Polosukhin},
  title        = {Attention is All you Need},
  booktitle    = {{NIPS}},
  pages        = {5998--6008},
  year         = {2017}
}

@inproceedings{DBLP:journals/corr/KingmaB14,
  author       = {Diederik P. Kingma and
                  Jimmy Ba},
  title        = {Adam: {A} Method for Stochastic Optimization},
  booktitle    = {{ICLR} (Poster)},
  year         = {2015}
}

@inproceedings{DBLP:journals/corr/SimonyanZ14a,
  author       = {Karen Simonyan and
                  Andrew Zisserman},
  title        = {Very Deep Convolutional Networks for Large-Scale Image Recognition},
  booktitle    = {{ICLR}},
  year         = {2015}
}

@inproceedings{DBLP:conf/iccv/CheferGW21,
  author       = {Hila Chefer and
                  Shir Gur and
                  Lior Wolf},
  title        = {Generic Attention-model Explainability for Interpreting Bi-Modal and
                  Encoder-Decoder Transformers},
  booktitle    = {{ICCV}},
  pages        = {387--396},
  publisher    = {{IEEE}},
  year         = {2021}
}

@inproceedings{DBLP:conf/icml/RadfordKHRGASAM21,
  author       = {Alec Radford and
                  Jong Wook Kim and
                  Chris Hallacy and
                  Aditya Ramesh and
                  Gabriel Goh and
                  Sandhini Agarwal and
                  Girish Sastry and
                  Amanda Askell and
                  Pamela Mishkin and
                  Jack Clark and
                  Gretchen Krueger and
                  Ilya Sutskever},
  title        = {Learning Transferable Visual Models From Natural Language Supervision},
  booktitle    = {{ICML}},
  series       = {Proceedings of Machine Learning Research},
  volume       = {139},
  pages        = {8748--8763},
  publisher    = {{PMLR}},
  year         = {2021}
}

@inproceedings{DBLP:conf/chi/ODonovanAH15,
  author       = {Peter O'Donovan and
                  Aseem Agarwala and
                  Aaron Hertzmann},
  title        = {DesignScape: Design with Interactive Layout Suggestions},
  booktitle    = {{CHI}},
  pages        = {1221--1224},
  publisher    = {{ACM}},
  year         = {2015}
}

@article{DBLP:journals/corr/abs-2110-06794,
  author       = {Mengxi Guo and
                  Dangqing Huang and
                  Xiaodong Xie},
  title        = {The Layout Generation Algorithm of Graphic Design Based on Transformer-CVAE},
  journal      = {CoRR},
  volume       = {abs/2110.06794},
  year         = {2021}
}

@inproceedings{DBLP:conf/chi/GuoJSLL0C21,
  author       = {Shunan Guo and
                  Zhuochen Jin and
                  Fuling Sun and
                  Jingwen Li and
                  Zhaorui Li and
                  Yang Shi and
                  Nan Cao},
  title        = {Vinci: An Intelligent Graphic Design System for Generating Advertising
                  Posters},
  booktitle    = {{CHI}},
  pages        = {577:1--577:17},
  publisher    = {{ACM}},
  year         = {2021}
}

@inproceedings{DBLP:conf/aaai/JiangSZL022,
  author       = {Zhaoyun Jiang and
                  Shizhao Sun and
                  Jihua Zhu and
                  Jian{-}Guang Lou and
                  Dongmei Zhang},
  title        = {Coarse-to-Fine Generative Modeling for Graphic Layouts},
  booktitle    = {{AAAI}},
  pages        = {1096--1103},
  publisher    = {{AAAI} Press},
  year         = {2022}
}

@article{DBLP:journals/tmm/LiZW22,
  author       = {Chenhui Li and
                  Peiying Zhang and
                  Changbo Wang},
  title        = {Harmonious Textual Layout Generation Over Natural Images via Deep
                  Aesthetics Learning},
  journal      = {{IEEE} Trans. Multim.},
  volume       = {24},
  pages        = {3416--3428},
  year         = {2022}
}

@article{DBLP:journals/tomccap/YangMXRL16,
  author       = {Xuyong Yang and
                  Tao Mei and
                  Ying{-}Qing Xu and
                  Yong Rui and
                  Shipeng Li},
  title        = {Automatic Generation of Visual-Textual Presentation Layout},
  journal      = {{ACM} Trans. Multim. Comput. Commun. Appl.},
  volume       = {12},
  number       = {2},
  pages        = {33:1--33:22},
  year         = {2016}
}

@inproceedings{DBLP:conf/icmcs/ZhangLW20,
  author       = {Peiying Zhang and
                  Chenhui Li and
                  Changbo Wang},
  title        = {Smarttext: Learning To Generate Harmonious Textual Layout Over Natural
                  Image},
  booktitle    = {{ICME}},
  pages        = {1--6},
  publisher    = {{IEEE}},
  year         = {2020}
}

@inproceedings{DBLP:conf/cvpr/KarrasLA19,
  author       = {Tero Karras and
                  Samuli Laine and
                  Timo Aila},
  title        = {A Style-Based Generator Architecture for Generative Adversarial Networks},
  booktitle    = {{CVPR}},
  pages        = {4401--4410},
  publisher    = {Computer Vision Foundation / {IEEE}},
  year         = {2019}
}

@inproceedings{DBLP:conf/cvpr/XuZGJX23,
  author       = {Chenchen Xu and
                  Min Zhou and
                  Tiezheng Ge and
                  Yuning Jiang and
                  Weiwei Xu},
  title        = {Unsupervised Domain Adaption with Pixel-Level Discriminator for Image-Aware
                  Layout Generation},
  booktitle    = {{CVPR}},
  pages        = {10114--10123},
  publisher    = {{IEEE}},
  year         = {2023}
}

@article{calic2007efficient,
  title={Efficient layout of comic-like video summaries},
  author={Calic, Janko and Gibson, David P and Campbell, Neill W},
  journal={IEEE Transactions on Circuits and Systems for Video Technology},
  volume={17},
  number={7},
  pages={931--936},
  year={2007},
  publisher={IEEE}
}

@article{cohn2013navigating,
  title={Navigating comics: An empirical and theoretical approach to strategies of reading comic page layouts},
  author={Cohn, Neil},
  journal={Frontiers in psychology},
  volume={4},
  pages={46474},
  year={2013},
  publisher={Frontiers}
}

@article{wang2021interactive,
  title={Interactive data comics},
  author={Wang, Zezhong and Romat, Hugo and Chevalier, Fanny and Riche, Nathalie Henry and Murray-Rust, Dave and Bach, Benjamin},
  journal={IEEE Transactions on Visualization and Computer Graphics},
  volume={28},
  number={1},
  pages={944--954},
  year={2021},
  publisher={IEEE}
}

@article{qiao2023design,
  title={Design Order Guided Visual Note Layout Optimization},
  author={Qiao, Xiaotian and Cao, Ying and Lau, Rynson WH},
  journal={IEEE Transactions on Visualization \& Computer Graphics},
  volume={29},
  number={09},
  pages={3922--3936},
  year={2023},
  publisher={IEEE Computer Society}
}

@article{DBLP:journals/tip/KanungoM03,
  title={Stochastic language models for style-directed layout analysis of document images},
  author={Kanungo, Tapas and Mao, Song},
  journal={IEEE Transactions on Image Processing},
  volume={12},
  number={5},
  pages={583--596},
  year={2003},
  publisher={IEEE}
}

@inproceedings{DBLP:conf/iui/SchrierDJWS08,
  title={Adaptive layout for dynamically aggregated documents},
  author={Schrier, Evan and Dontcheva, Mira and Jacobs, Charles and Wade, Geraldine and Salesin, David},
  booktitle={Proceedings of the 13th international conference on Intelligent user interfaces},
  pages={99--108},
  year={2008}
}

@article{DBLP:journals/tip/HedjamNKC15,
  title={Influence of color-to-gray conversion on the performance of document image binarization: Toward a novel optimization problem},
  author={Hedjam, Rachid and Nafchi, Hossein Ziaei and Kalacska, Margaret and Cheriet, Mohamed},
  journal={IEEE transactions on image processing},
  volume={24},
  number={11},
  pages={3637--3651},
  year={2015},
  publisher={IEEE}
}

@inproceedings{qiang2016learning,
  title={Learning to generate posters of scientific papers},
  author={Qiang, Yuting and Fu, Yanwei and Guo, Yanwen and Zhou, Zhi-Hua and Sigal, Leonid},
  booktitle={Proceedings of the AAAI Conference on Artificial Intelligence},
  volume={30},
  number={1},
  year={2016}
}

@article{qiang2019learning,
  title={Learning to generate posters of scientific papers by probabilistic graphical models},
  author={Qiang, Yu-Ting and Fu, Yan-Wei and Yu, Xiao and Guo, Yan-Wen and Zhou, Zhi-Hua and Sigal, Leonid},
  journal={Journal of Computer Science and Technology},
  volume={34},
  pages={155--169},
  year={2019},
  publisher={Springer}
}

@article{DBLP:journals/jzusc/YouJYYS20,
  title={Automatic synthesis of advertising images according to a specified style},
  author={You, Wei-tao and Jiang, Hao and Yang, Zhi-yuan and Yang, Chang-yuan and Sun, Ling-yun},
  journal={Frontiers of Information Technology \& Electronic Engineering},
  volume={21},
  number={10},
  pages={1455--1466},
  year={2020},
  publisher={Springer}
}

@article{DBLP:journals/tog/PangCLC16,
  title={Directing user attention via visual flow on web designs},
  author={Pang, Xufang and Cao, Ying and Lau, Rynson WH and Chan, Antoni B},
  journal={ACM Transactions on Graphics (TOG)},
  volume={35},
  number={6},
  pages={1--11},
  year={2016},
  publisher={ACM New York, NY, USA}
}

@inproceedings{DBLP:conf/mm/ZhangHRYXH17,
  title={Layout style modeling for automating banner design},
  author={Zhang, Yunke and Hu, Kangkang and Ren, Peiran and Yang, Changyuan and Xu, Weiwei and Hua, Xian-Sheng},
  booktitle={Proceedings of the on Thematic Workshops of ACM Multimedia 2017},
  pages={451--459},
  year={2017}
}

@inproceedings{DBLP:journals/india,
  title={Enabling hyper-personalisation: Automated ad creative generation and ranking for fashion e-commerce},
  author={Vempati, Sreekanth and Malayil, Korah T and Sruthi, V and Sandeep, R},
  booktitle={Fashion Recommender Systems},
  pages={25--48},
  year={2020},
  organization={Springer}
}

@inproceedings{jiang2023layoutformer++,
  title={Layoutformer++: Conditional graphic layout generation via constraint serialization and decoding space restriction},
  author={Jiang, Zhaoyun and Guo, Jiaqi and Sun, Shizhao and Deng, Huayu and Wu, Zhongkai and Mijovic, Vuksan and Yang, Zijiang James and Lou, Jian-Guang and Zhang, Dongmei},
  booktitle={Proceedings of the IEEE/CVF Conference on Computer Vision and Pattern Recognition},
  pages={18403--18412},
  year={2023}
}

@inproceedings{inoue2023layoutdm,
  title={Layoutdm: Discrete diffusion model for controllable layout generation},
  author={Inoue, Naoto and Kikuchi, Kotaro and Simo-Serra, Edgar and Otani, Mayu and Yamaguchi, Kota},
  booktitle={Proceedings of the IEEE/CVF Conference on Computer Vision and Pattern Recognition},
  pages={10167--10176},
  year={2023}
}

@inproceedings{levi2023dlt,
  title={Dlt: Conditioned layout generation with joint discrete-continuous diffusion layout transformer},
  author={Levi, Elad and Brosh, Eli and Mykhailych, Mykola and Perez, Meir},
  booktitle={Proceedings of the IEEE/CVF International Conference on Computer Vision},
  pages={2106--2115},
  year={2023}
}

@inproceedings{li2023relation,
  title={Relation-aware diffusion model for controllable poster layout generation},
  author={Li, Fengheng and Liu, An and Feng, Wei and Zhu, Honghe and Li, Yaoyu and Zhang, Zheng and Lv, Jingjing and Zhu, Xin and Shen, Junjie and Lin, Zhangang and others},
  booktitle={Proceedings of the 32nd ACM International Conference on Information and Knowledge Management},
  pages={1249--1258},
  year={2023}
}

@inproceedings{horita2024retrieval,
  title={Retrieval-Augmented Layout Transformer for Content-Aware Layout Generation},
  author={Horita, Daichi and Inoue, Naoto and Kikuchi, Kotaro and Yamaguchi, Kota and Aizawa, Kiyoharu},
  booktitle={Proceedings of the IEEE/CVF Conference on Computer Vision and Pattern Recognition},
  pages={67--76},
  year={2024}
}

@inproceedings{kong2022blt,
  title={Blt: Bidirectional layout transformer for controllable layout generation},
  author={Kong, Xiang and Jiang, Lu and Chang, Huiwen and Zhang, Han and Hao, Yuan and Gong, Haifeng and Essa, Irfan},
  booktitle={European Conference on Computer Vision},
  pages={474--490},
  year={2022},
  organization={Springer}
}

@article{yu2022layoutdetr,
  title={LayoutDETR: detection transformer is a good multimodal layout designer},
  author={Yu, Ning and Chen, Chia-Chih and Chen, Zeyuan and Meng, Rui and Wu, Gang and Josel, Paul and Niebles, Juan Carlos and Xiong, Caiming and Xu, Ran},
  journal={arXiv preprint arXiv:2212.09877},
  year={2022}
}

@inproceedings{xie2021canvasemb,
  title={Canvasemb: Learning layout representation with large-scale pre-training for graphic design},
  author={Xie, Yuxi and Huang, Danqing and Wang, Jinpeng and Lin, Chin-Yew},
  booktitle={Proceedings of the 29th ACM international conference on multimedia},
  pages={4100--4108},
  year={2021}
}

@article{xuan2023cvae,
  title={CVAE-LAYOUT: automatic furniture layout with constraints},
  author={Xuan, Yixin and Song, Chao and Jin, Jianqiu and Yang, Bailin},
  journal={The Visual Computer},
  pages={1--15},
  year={2023},
  publisher={Springer}
}

@article{fan2023real,
  title={Real-scene-constrained virtual scene layout synthesis for mixed reality},
  author={Fan, Runze and Wang, Lili and Liu, Xinda and Im, Sio Kei and Lam, Chan Tong},
  journal={The Visual Computer},
  pages={1--21},
  year={2023},
  publisher={Springer}
}

@article{liang2023sketch2wireframe,
  title={Sketch2Wireframe: an automatic framework for transforming hand-drawn sketches to digital wireframes in UI design},
  author={Liang, Xudong and Lin, Tao},
  journal={The Visual Computer},
  pages={1--11},
  year={2023},
  publisher={Springer}
}

% \newpage

 

% \end{IEEEbiography}

\vspace{11pt}

\vfill

\end{document}

\title{Image-aware Layout Generation\\ with User Constraints for Poster Design \\ Supplementary Material}

\author{Chenchen Xu, Min Zhou, Tiezheng Ge, and Weiwei Xu
        % <-this % stops a space
\thanks{Chenchen Xu and Weiwei Xu are with the State Key Lab of CAD\&CG, Zhejiang University, China. Min Zhou and Tiezheng Ge are with the Alibaba Group, China.}
\thanks{E-mail: xuchenchen@zju.edu.cn, yunqi.zm@alibaba-inc.com, tiezheng.gtz@alibaba-inc.com, and xww@cad.zju.edu.cn}
}

\markboth{Supplementary Material}%
{Shell \MakeLowercase{\textit{et al.}}: A Sample Article Using IEEEtran.cls for IEEE Journals}


\maketitle

\begin{figure*}[h!]
    \centering
    \hspace{0cm}\includegraphics[width=\textwidth]{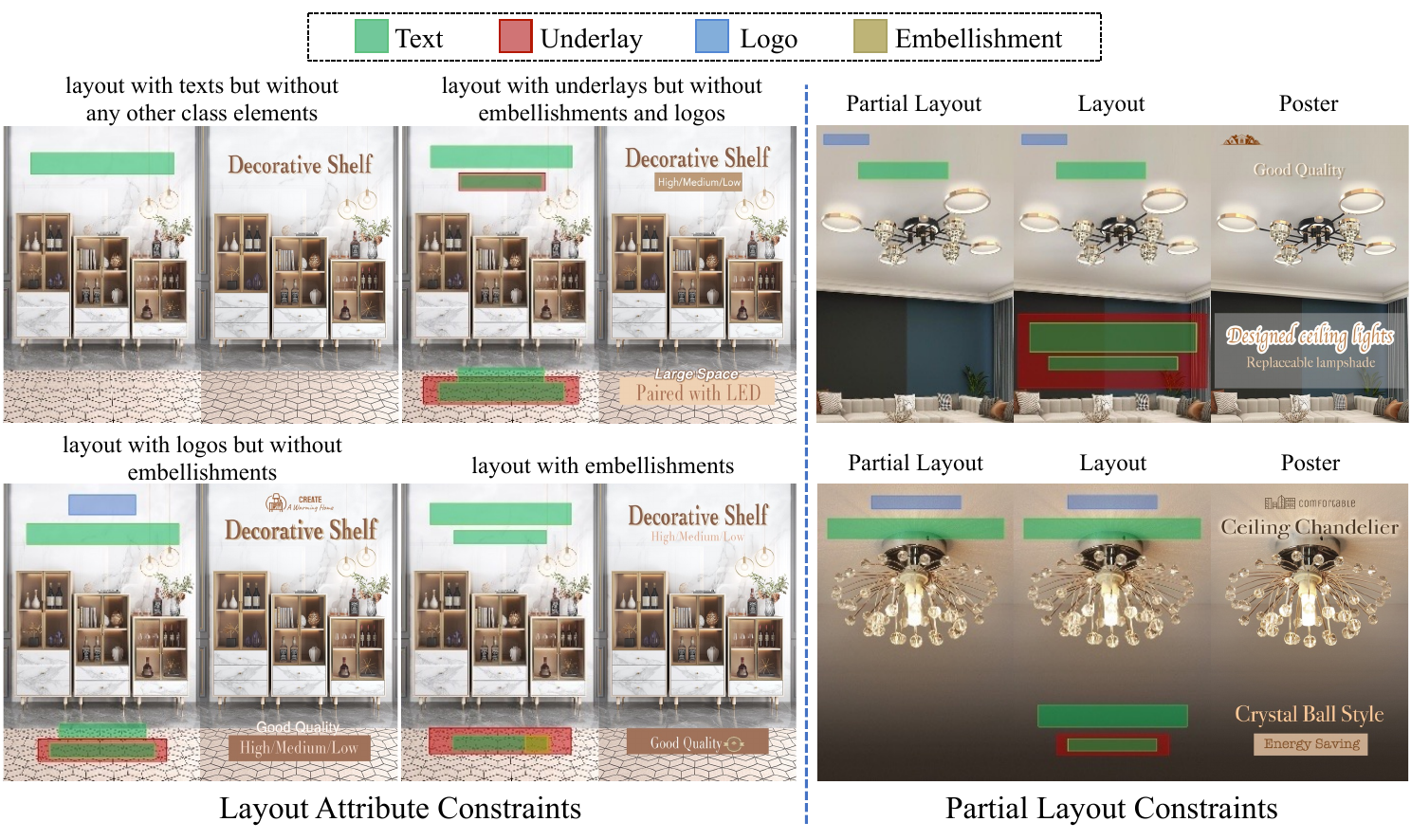}
    \caption{{\bf Examples of generated layouts and posters with image contents and user constraints.} Our model generates image-aware layouts that adhere to layout attribute constraints (left) and partial layout constraints (right), which can be used to generate advertising posters.}
    \label{fig:introduction}
\end{figure*}

\section{Poster Design with Our Layout Results}
\begin{figure*}[]
    \centering
    \hspace{0cm}\includegraphics[width=\textwidth]{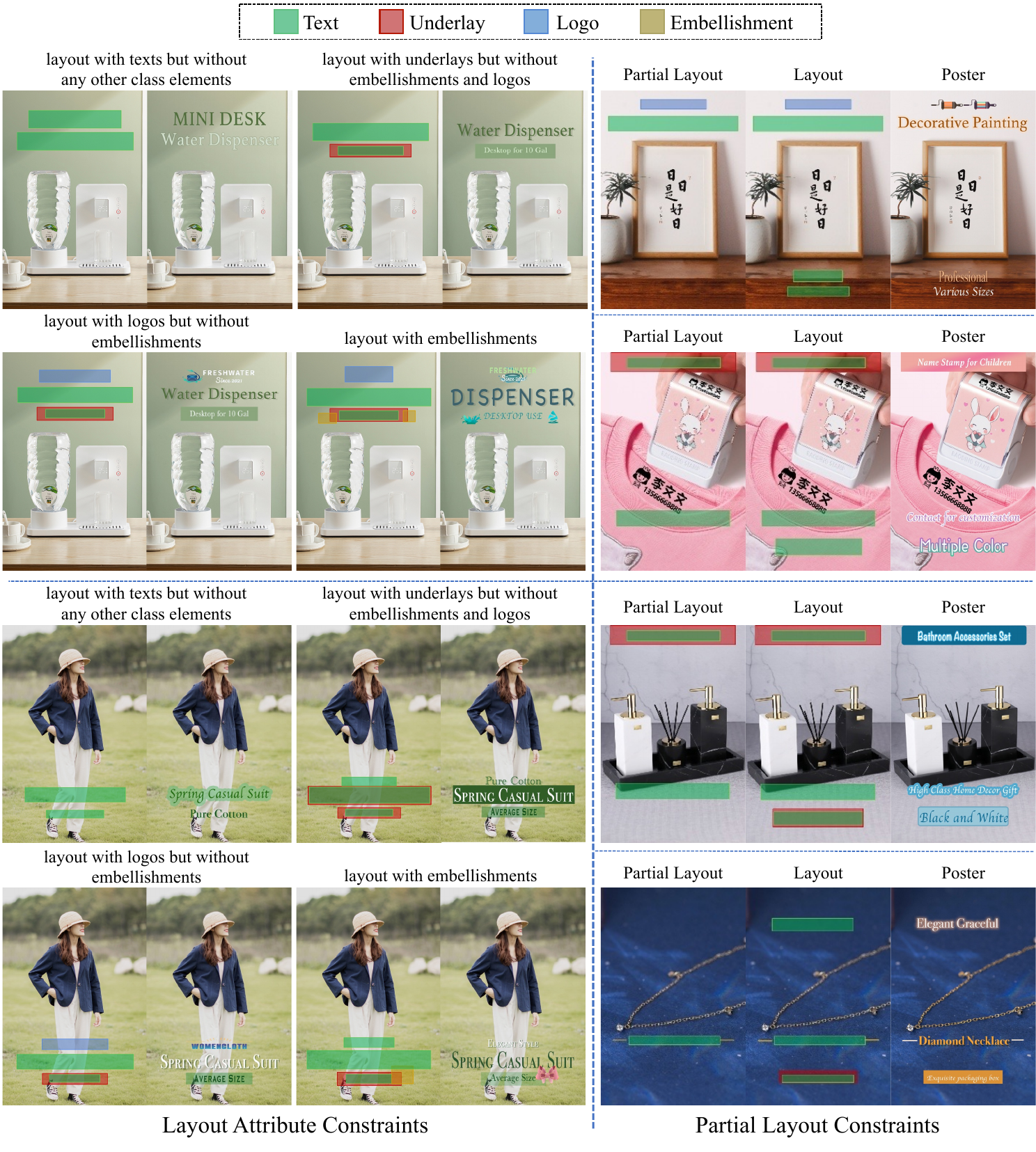}
    \caption{Demonstration of advertising posters based on graphic layouts generated by IUC-Layout network conditioned on product images and various user constraints, including layout attributes (left) and partial layouts (right).}
    \label{fig:poster}
\end{figure*}

\IEEEPARstart{D}{esigners} have applied graphic layouts generated by our IUC-Layout network to design aesthetic advertising posters. As shown in Fig.~\ref{fig:introduction}, our model generates graphic layouts adhering to diverse user constraints, including layout attributes and partial layouts. These generated graphic layouts serve as resources for designers or automatic rendering systems to produce aesthetically pleasing advertising posters. More demonstrations of advertising poster design with our layout results are shown in Fig.~\ref{fig:poster} and the videos, which are provided in our supplementary materials. 

The left part of Fig.~\ref{fig:poster} demonstrates our model's capability to generate corresponding style image-aware layouts that satisfy various attribute constraints. These layouts provide a foundation for creating posters that meet various display requirements of products. For instance, if the attribute constraint is "layout with logo but without embellishment", the generated layout will include logo elements,  enabling the designed poster to effectively showcase the product's logo information. The right part of Fig.~\ref{fig:poster} displays layouts generated by the model with the introduced partial-constraint loss. These layouts exhibit precise alignment with the provided partial layout, illustrating the model's proficiency in supplementing incomplete layouts.

% In the first column in \ref{fig:attribute-poster}, when the layout attribute is "layout with text but without any other type elements", generated layouts consist of text elements only. Similarly, the 2nd, 3rd, and 4th columns demonstrate that our model can also generate corresponding style image-aware layouts to satisfy other attribute constraints. These layouts can serve as the foundation for designing posters that meet various display requirements of products. For instance, if the attribute requirement is "layout with logo but without embellishment", the generated layout will include logo elements, and the designed poster can display the product's logo information.

\section{Reason of Summarizing Layout Attribute Constraints into Four Types}\label{attribute_FT}
We conducted a statistical analysis of 54,546 posters, comprising 162,647 text elements (61.12\%), 60,556 underlay elements (22.76\%), 34,303 logo elements (12.89\%), and 8,598 embellishment elements (3.23\%). Considering posters' characteristics, user requirements, layout styles, and elements' relationships, we define four attribute types: (1) layout with texts but without any other class elements, (2) layout with underlays but without embellishments and logos, (3) layout with logos but without embellishments, and (4) layout with embellishments. In practical application, layout attributes can exhibit significant variations based on the dataset and the specific layout application scenarios. These four layout attributes provide inspiration and case studies for attribute-controlled layout generation research.

\begin{table}[!t]
    \centering
    \setlength{\tabcolsep}{0.88mm}
    \renewcommand{\arraystretch}{1.5}
    \caption{Five sets of four-channel Gaussian noise with varying means based on different layout attributes.}
    {
    \scalebox{1.1}{
    \begin{tabular}{c|c}
    \hline
         \bf{layout attribute}  &\bf{means of four-channel noise}\\
    \hline     
         layout with texts \\ but without any other class elements &$(1, -1, -1, 1)$\\
    \hline     
         layout with underlays \\ but without embellishments and logos &$(1, -1, 1, -1)$ \\
    \hline
         layout with logos \\ but without embellishments &$(1, 1, -1, -1)$\\
    \hline
         layout with embellishments &$(1, 1, 1, 1)$\\
    \hline
         unspecified layout attribute &$(0, 0, 0, 0)$\\
    \hline
    \end{tabular}
    }
    }
    \label{tab:noise discription}
\end{table}

\section{Dataset Split}
Following~\cite{DBLP:conf/ijcai/ZhouXMGJX22}, we used 54,546 annotated posters for training and 1,000 clean product images for testing. The annotated information and analyses conducted in Sec.~\ref{attribute_FT} guided the division of these annotated samples into four attribute training sets for the model. During the model training process with partial layout constraints, 25\% of the annotated bounding box information within each layout was randomly selected to form the partial layout. In testing, we assessed the effectiveness of five types of constraints, as detailed in Tab.~\ref{tab:noise discription}, for each sample in terms of attribute constraints. Regarding partial layout constraints, we introduced randomly generated partial layouts to serve as constraints.

\section{More Evaluations}
In response to the paper's quantitative evaluations, this section supplements the metrics' explanation and provides more qualitative comparisons to further demonstrate the effectiveness of IUC-Layout network.

\subsection{Metrics}
We follow \cite{DBLP:conf/ijcai/ZhouXMGJX22,DBLP:conf/cvpr/XuZGJX23} to utilize composition-relevant and graphic metrics to evaluate the performance of our model. Composition-relevant metrics include $R_{com}$, $R_{shm}$ \cite{DBLP:journals/corr/SimonyanZ14a}, and $R_{sub}$ \cite{DBLP:conf/iccv/CheferGW21,DBLP:conf/icml/RadfordKHRGASAM21}, which measure background complexity, subject occlusion, and product occlusion, respectively. Graphic metrics consist of layout overlap $R_{ove}$ \cite{DBLP:journals/pami/LiYHZX21,DBLP:journals/tvcg/LiY0LWX21}, underlay overlap $R_{und}$ \cite{DBLP:conf/ijcai/ZhouXMGJX22,DBLP:conf/cvpr/XuZGJX23}, layout alignment $R_{ali}$ \cite{DBLP:journals/pami/LiYHZX21,DBLP:journals/tvcg/LiY0LWX21} and the ratio of non-empty layouts $R_{occ}$. To evaluate the model's performance on layout attribute and partial layout constraints, we introduce metrics of $R_{lac}$ and $R_{plc}$. $R_{lac}$ indicates the ratio of generated layouts that conform to the given attribute constraints. $R_{plc}$ is used to quantify the average difference between given partial layout constraints and generated layouts, and it can be formulated as:
\begin{equation}
    R_{plc} = \frac{1}{\mathcal{N}} \sum^{\mathcal{N}}_{i=1}
    \left\vert
    Pred_{i-id} - PL_{i-id}
    \right\vert
\end{equation}
$\mathcal{N}$ represents the total number of given partial layout information. $Pred_{i\mbox{-}id}$ represents the predicted value at the $i\mbox{-}th$ index $id$ of model output according to given partial layout index $id$. $PL_{i\mbox{-}id}$ is the value of given partial layout at the $i\mbox{-}th$ index. By combining these metrics, we can assess the model's performance in terms of graphic quality, product content relevance, layout attribute constraint, and partial layout consistency.

\subsection{Layout Attribute Constraints}
\begin{figure*}[!t]
    \centering
    \hspace{0cm}\includegraphics[width=\textwidth]{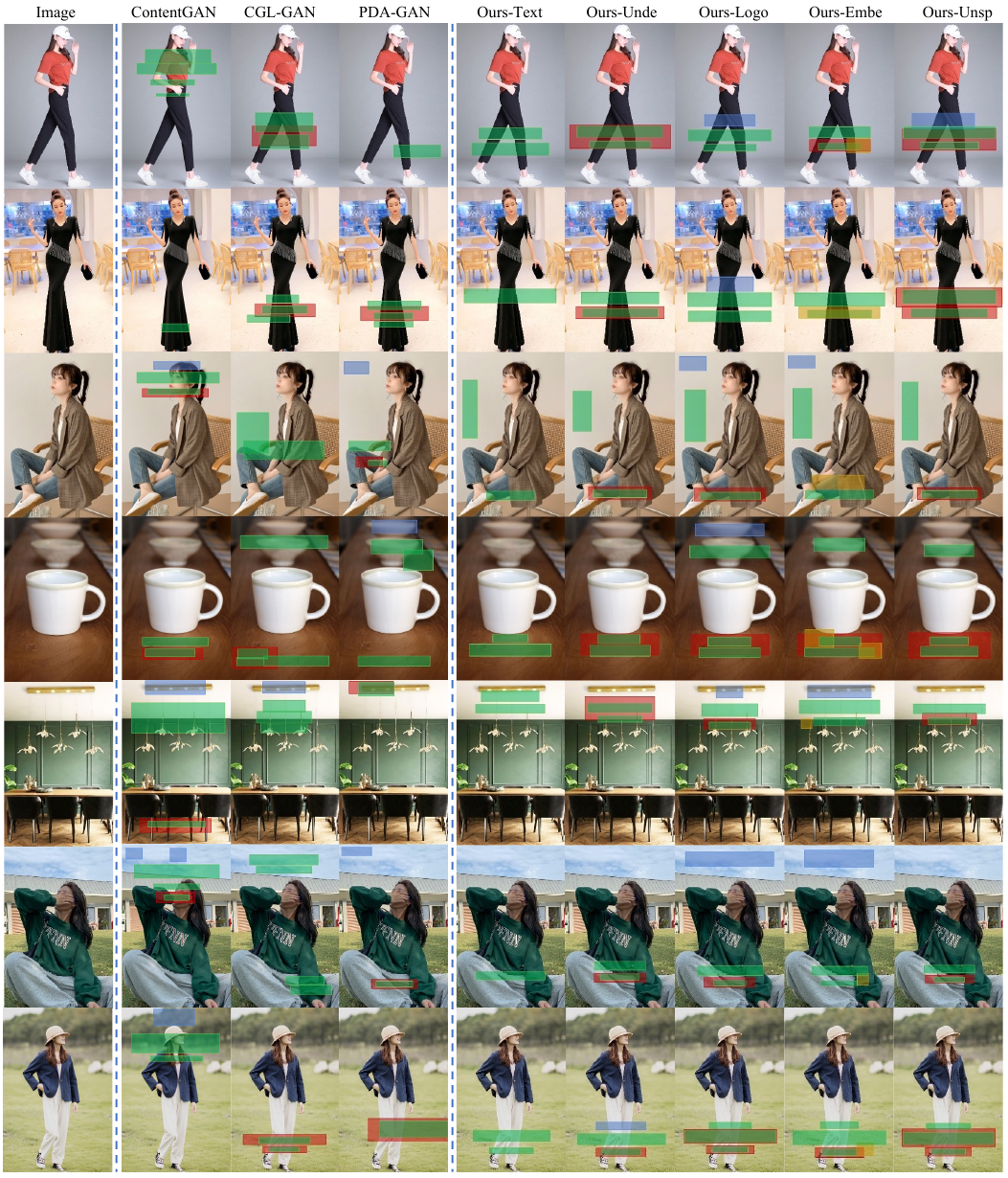}
    \caption{{\bf Qualitative evaluation with image-aware models.} The layouts in each row are conditioned on the same product image, while the ones in a column are generated by the same model. $Ours\mbox{-}Text$ as a sample means our model with the attribute of "layout with texts but without any other class elements".}
    \label{fig:image-aware}
\end{figure*}

\noindent\textbf{Comparison with image-aware methods.} 
We first compare our model with image-aware methods \cite{DBLP:journals/tog/ZhengQCL19,DBLP:conf/ijcai/ZhouXMGJX22,DBLP:conf/cvpr/XuZGJX23} in Fig.~\ref{fig:image-aware}. These comparisons further demonstrate that our model can generate different layouts according to various attributes for the same product image, showcasing its ability to express layout attributes. For instance, the first-second product image in Fig.~\ref{fig:image-aware} required displaying the product's logo, previous models were unable to meet such demand. By specifying "layout with logo but without embellishment" as the attribute for IUC-Layout network, the generated layout can effectively present the logo information.

In addition, as evident from the 2nd, 3rd, and 4th columns of Fig.~\ref{fig:image-aware}, layouts generated by previous works exhibit very few or even no embellishment elements due to the rarity of such elements in the CGL-Dataset \cite{DBLP:conf/ijcai/ZhouXMGJX22}. In contrast, our model tends to generate embellishment elements to decorate the graphic layout when "layout with embellishment" is the specified attribute, as demonstrated in the 8th column. These evaluations further underscore the effectiveness of IUC-Layout network with the proposed attribute-consistent loss and attribute-disentangled loss in expressing various layout attributes.

\begin{figure*}[!t]
    \centering
    \hspace{0cm}\includegraphics[width=\textwidth]{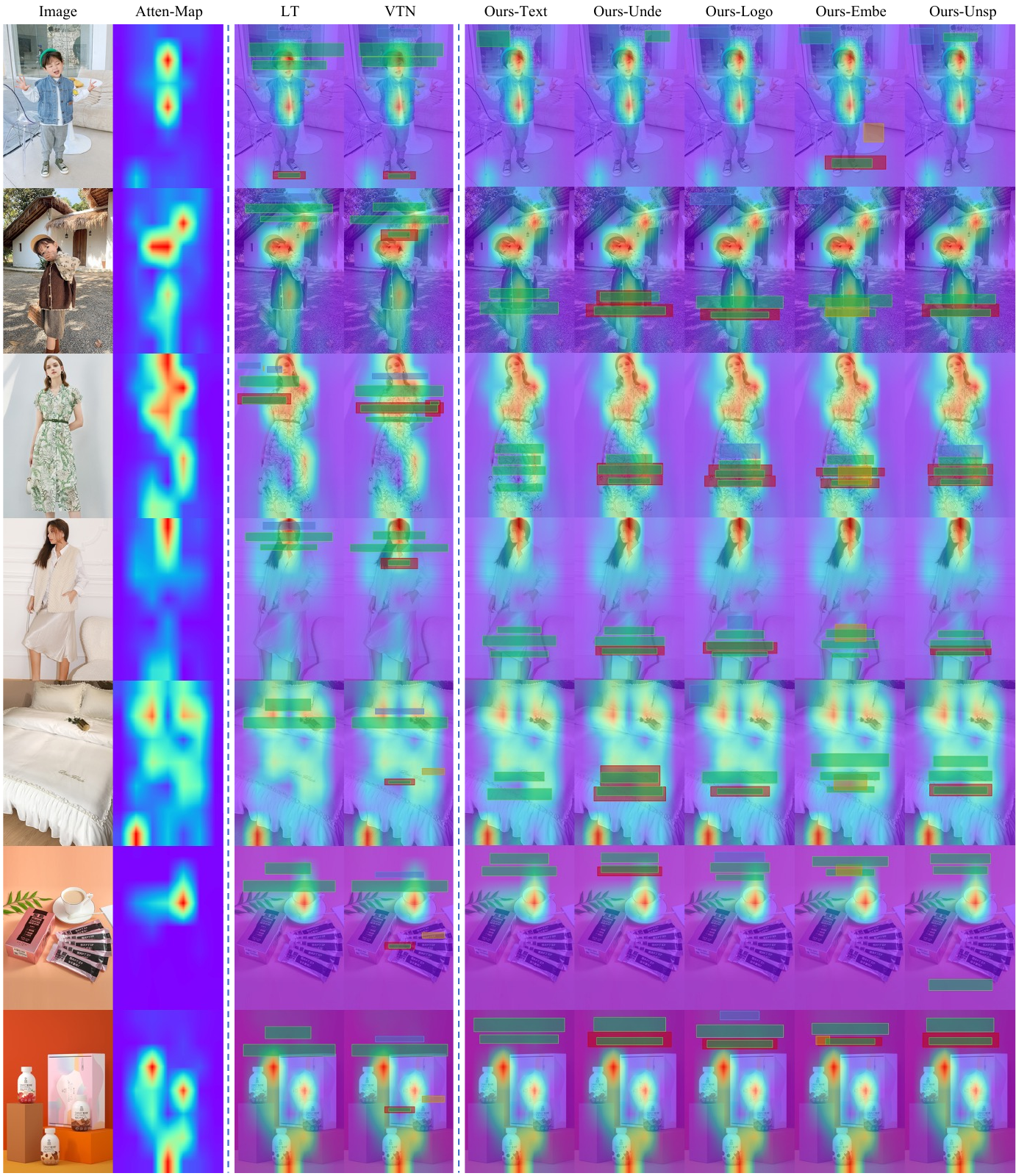}
    \caption{{\bf Qualitative evaluation with image-agnostic models.} Layouts in each row are conditioned on the same image with product attention map $Atten\mbox{-}Map$ \cite{DBLP:conf/iccv/CheferGW21,DBLP:conf/icml/RadfordKHRGASAM21}. Those layouts in a column are generated from the same model. $LT$ and $VTN$ represent LayoutTransformer and LayoutVTN, respectively.}
    \label{fig:image-agnostic}
\end{figure*}

\noindent\textbf{Comparison with image-agnostic methods.} We also present additional comparisons with image-agnostic models, as depicted in Fig.~\ref{fig:image-agnostic}. The 3rd and 4th columns illustrate that layouts generated by LayoutTransformer \cite{DBLP:conf/iccv/GuptaLA0MS21} and LayoutVTN \cite{DBLP:conf/cvpr/ArroyoPT21} randomly obscure subject areas, potentially diminishing the effective presentation of product information. In contrast, layouts generated by IUC-Layout network under various attribute constraints tend to occlude regions with lower product attention. Moreover, graphic layouts generated by our model can effectively avoid the critical region of subjects, such as the human head and face, thereby improving the visual aesthetics of the posters designed from these layouts. Therefore, these comparisons demonstrate that IUC-Layout can not only express different attributes but also understand the content of product images.

\subsection{Partial Layout Constraints}
\begin{figure*}[!t]
    \centering
    \hspace{0cm}\includegraphics[width=\textwidth]{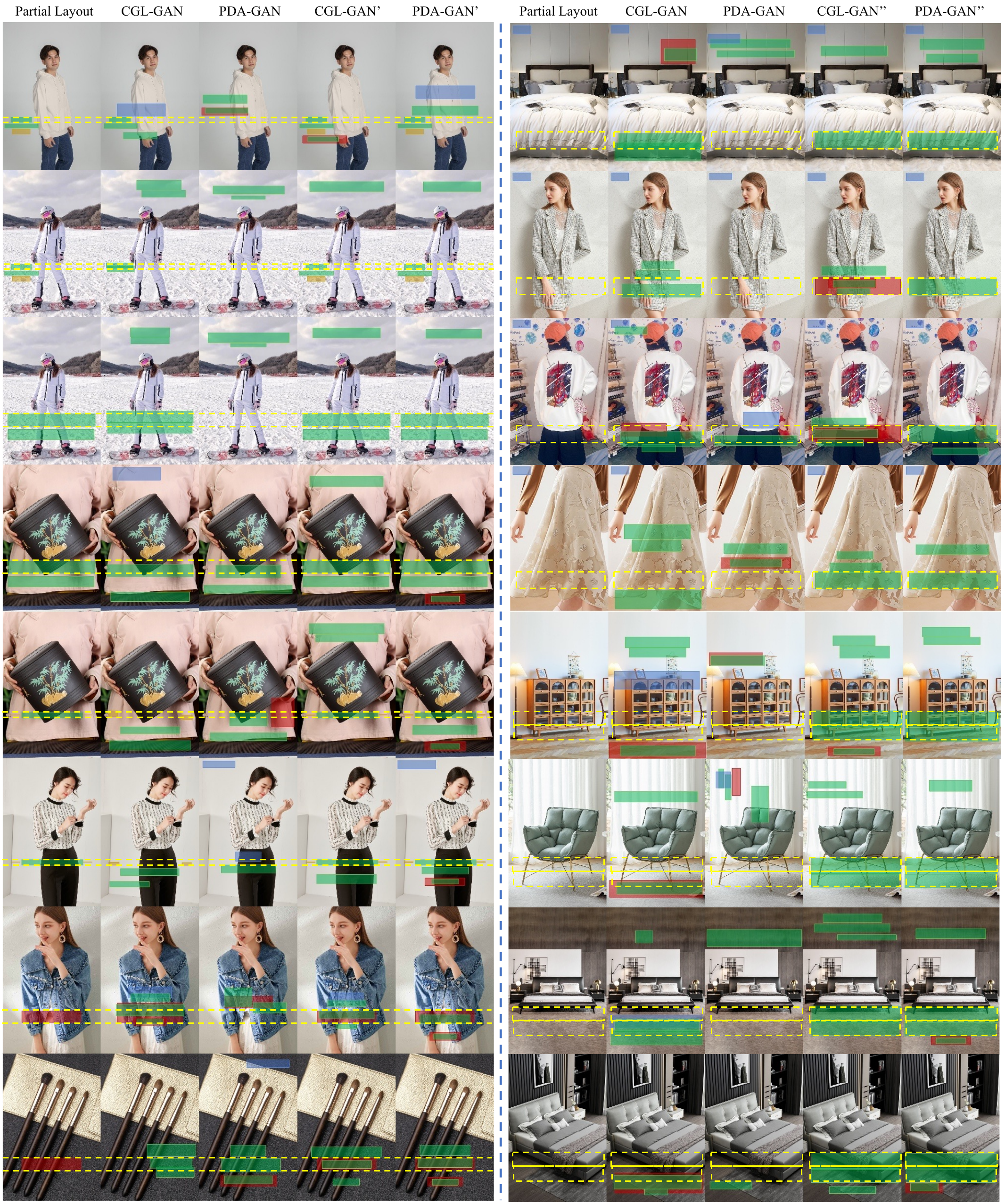}
    \caption{Partial layouts on the left encompass complete elements, while the right side contains incomplete element information. The symbol $'$ (or $''$) indicates the model introducing $L_{P}$ (or $L_{PL_{rm}}$). The yellow dashed line is used to measure the alignment between the generated layouts and the given partial layout.}
    \label{fig:partial-layout}
\end{figure*}

\noindent\textbf{Partial layout consists of complete element information.}
The partial layout in the left part of Fig.~\ref{fig:partial-layout} consists of complete elements, including the class and position information. Ablation experiments on previous methods show that applying $L_{P}$ to these models significantly improves the consistency between the given partial layout and the generated complete layouts. Interestingly, as shown in the last two rows, when a partial layout is given an underlay element, the model can correspondingly generate a text element, resulting in a coordinated layout.

\noindent\textbf{Partial layout with incomplete element information.}
In the right part, the first four partial layouts include a logo element and coordinates of another element without class information. The last four partial layouts provide only coordinates of two boxes without class information. The 2nd and 3rd columns demonstrate that previous models fail to handle the partial layout constraint with incomplete element information, such as coordinates only. In contrast, as shown in the last two columns, these models trained with $L_{PL_{rm}}$ perform well in the task of partial layout constraints with incomplete element information. Specifically, when given position information only, the corresponding class of the generated element can be text or other classes.

The above qualitative comparisons further demonstrate that the proposed $L_{P}$ and random mask can be readily applied to other models to fulfill partial layout constraints.

\section{Expand to other applications.}
Our principles extend beyond poster generation to other domains:
1) Other areas of graphic design: Web pages, magazines, and newspapers can also benefit from our principles. For instance, integrating the partial-constraint loss into models for these tasks, similar to the application on CGL-GAN~\cite{DBLP:conf/ijcai/ZhouXMGJX22} and PDA-GAN~\cite{DBLP:conf/cvpr/XuZGJX23} in the experiment section, ensures that generated layouts meet user requirements.
2) Film design and other artistic domains: Our principles are applicable to designing and arranging segments, titles, subtitles, and other visual elements in film and video artistry. This facilitates generating multiple layouts aligning with the script's requirements, offering directors options to enhance production efficiency and artistic vision, ultimately creating captivating cinematic experiences.
In summary, our principles streamline layout generation across domains, empowering designers to create tailored visual content efficiently.

\section{Ethical and societal impact.}
\noindent\textbf{Ethical implications.} 
This paper aims at creating image-aware layouts with user constraints to assist designers in crafting aesthetic posters. Our model generates graphic layouts consisting solely of bounding boxes and without any discriminatory or harmful content, thus complying with ethical standards. 

\noindent \textbf{Societal impact.} 
This research holds significant societal implications across multiple dimensions. Firstly, it plays a pivotal role in breaking down design barriers by serving as a model that not only provides creative ideas but also accelerates layout generation, enhancing efficiency for designers. Moreover, it addresses the issue of limited design experience by empowering individuals to successfully create layouts for advertising posters. Another noteworthy aspect is its impact on experienced designers, fueling creativity through the encouragement of experimentation with various layout constraints for more appealing posters. This enhanced advertising quality contributes significantly to brand awareness and commercial value. Finally, the model emerges as a valuable tool for design education and training, aiding students and professionals in understanding layout design principles and techniques. Therefore, this work makes significant contributions to various dimensions of societal development.

% \begin{figure*}[!t]
%     \centering
%     \hspace{0cm}\includegraphics[width=\textwidth]{pictures/SM4.pdf}
%     \caption{Demonstration of advertising posters based on graphic layouts generated by IUC-Layout conditioned on product images and various attributes.}
%     \label{fig:attribute-poster}
% \end{figure*}

% \begin{figure*}[!t]
%     \centering
%     \hspace{0cm}\includegraphics[width=\textwidth]{pictures/SM6.pdf}
%     \caption{Demonstration of posters based on graphic layouts generated by previous models with proposed $L_{PL}$ ($'$) and $L_{PL_{rm}}$ ($''$).}
%     \label{fig:partial-1}
% \end{figure*}

% \begin{figure*}[!t]
%     \centering
%     \hspace{0cm}\includegraphics[width=\textwidth]{pictures/SM5.pdf}
%     \caption{Demonstration of posters based on graphic layouts generated by previous models with proposed $L_{PL}$ ($'$) and $L_{PL_{rm}}$ ($''$).}
%     \label{fig:partial-2}
% \end{figure*}

% \section*{Acknowledgments}
% This should be a simple paragraph before the References to thank those individuals and institutions who have supported your work on this article.

\bibliographystyle{IEEEtran}
\bibliography{cite}

% \newpage

 

% \end{IEEEbiography}

\vspace{11pt}

\vfill